\documentclass{article}

\usepackage{arxiv}

\usepackage{amsmath}        % math symbols
\usepackage{float}          % for using H option for figure placement
\usepackage[utf8]{inputenc} % allow utf-8 input
\usepackage[T1]{fontenc}    % use 8-bit T1 fonts
\usepackage{hyperref}       % hyperlinks
\usepackage{url}            % simple URL typesetting
\usepackage{booktabs}       % professional-quality tables
\usepackage{amsfonts}       % blackboard math symbols
\usepackage{nicefrac}       % compact symbols for 1/2, etc.
\usepackage{microtype}      % microtypography
\usepackage{lipsum}
\usepackage{graphicx}
\graphicspath{ {./images/} }

\title{Causal emergence is widespread across measures of causation}

\author{
  Renzo Comolatti \\
  University of Milan\\
  Milan, MI, Italy \\
  \texttt{renzo.com@gmail.com} \\
   \And
  Erik Hoel\thanks{Corresponding author} \\
  Allen Discovery Center\\
  Tufts University\\
  Medford, MA, USA \\
  \texttt{erik.hoel@tufts.edu} \\
}

\begin{document}
\maketitle
\begin{abstract}
Causal emergence is the theory that macroscales can reduce the noise in causal relationships, leading to stronger causes at the macroscale. First identified using the effective information and later the integrated information in model systems, causal emergence has been analyzed in real data across the sciences since. But is it simply a quirk of these original measures? To answer this question we examined over a dozen popular measures of causation, all independently developed and widely used, and spanning different fields from philosophy to statistics to psychology to genetics. All showed cases of causal emergence. This is because, we prove, measures of causation are based on a small set of related "causal primitives." This consilience of independently-developed measures of causation shows that macroscale causation is a general fact about causal relationships, is scientifically detectable, and is not a quirk of any particular measure of causation. This finding sets the science of emergence on firmer ground, opening the door for the detection of intrinsic scales of function in complex systems, as well as assisting with scientific modeling and experimental interventions.
\end{abstract}

% keywords can be removed
%\keywords{First keyword \and Second keyword \and More}

\section{Introduction}
While causation has historically been a subject of philosophical debate, work over the last few decades has shown that metaphysical speculations can be put aside in favor of mathematical formalisms \cite{pearl_causality_2009}. Indeed, causation is referenced universally throughout the sciences without metaphysical commitments, and mathematical treatments of causation come from diverse scientific fields like psychology and statistics \cite{fitelson_probabilistic_2010}. E.g., in the neurosciences, people have used a number of measures of causation to track the result of experimental interventions \cite{massimini_perturbational_2009, chettih_single-neuron_2019, sporns_brain_2007, fingelkurts_functional_2005}. However, due to this plethora of measures of causation, one might argue there is subjectivity in terms of what counts as a cause or not, since a particular scientist might prefer one measure over another.

Here we offer a way around this problem by showing that popular measures of causation are mathematically related, behave very similarly under many conditions, and are sensitive to the same fundamental properties. Indeed, all the measures we examined turned out to be based on a small set of what we dub \textit{causal primitives}. By showing how over a dozen measures of causation are grounded in the same primitives, we reveal there is widespread consilience in terms of what constitutes a strong or weak cause (or more generally, a strong or weak causal relationship). This research obviates the need to arrive at a lone measure of causation that researchers must universally agree upon, but rather reveals a sphere of viable measures with significant overlap (much like the definitions of "complexity" in complex systems science \cite{gell-mann_what_1995}). By focusing on the agreement between a family of well-accepted and closely-related measures, we can move on to understanding other causal phenomena.

One such important phenomena is causal emergence, which is when a causal relationship is stronger at the macroscale \cite{hoel_quantifying_2013}. While at first counterintuitive, causal emergence is grounded in the fact that macroscales can lead to noise reduction in causal relationships. Broadly, this noise is synonymous with uncertainty, which can come from different sources, and macroscale models can reduce or minimize this error. In such cases, universal reduction is unworkable, since such reduction would "leave some causation on the table," even though the macroscale supervenes (is fixed by) its underlying microscale. Note that claims of emergence are not metaphysical speculations. They have real consequences. For example, emergent macroscale models are more useful to intervene on and understand the system in question with \cite{chvykov_causal_2020}; causal emergence can reveal the intrinsic scales of function in opaque non-engineered systems where the scale of interest is unknown, like in gene regulatory networks \cite{hoel_emergence_2020}; it can also be used to find partitions of directed graphs and is more common in biological networks vs. technological networks \cite{klein_emergence_2020}; it has revealed novel groupings of cellular automata rules \cite{varley_causal_2020}; causal emergence has been used to identify macrostates in timeseries data using artificial neural networks \cite{zhang_neural_2022}; there's even some evidence that evolution selects for causal emergence, possibly because macroscales that are causally-emergent have been shown to be more robust to knock-outs and attacks \cite{klein_evolution_2021}. Such questions are relevant across the sciences, e.g., there are fundamental questions about what scale is of most importance in brain function \cite{yuste_neuron_2015, buxhoeveden_minicolumn_2002, yeo_organization_2011} that only a scientific theory of emergence can resolve; indeed, causal emergence might explain the spatiotemporal scale of consciousness in the brain \cite{hoel_can_2016, chang_information_2020}.

However, evidence for causal emergence has previously been confined to a small set of measures: first, the effective information \cite{hoel_quantifying_2013, hoel_when_2017, klein_emergence_2020}, and then later, the integrated information \cite{hoel_can_2016, marshall_black-boxing_2018}. Both these measures, grounded in information theory, are designed to capture subtly different aspects of causation. Yet they are related mathematically and involve similar background assumptions. Because of this, some have criticized the results of the measures, pointing to how interventions are performed (e.g., perhaps effective information requiring a maximum-entropy intervention distribution means it's somehow invalid or assumptive \cite{aaronson_higher-level_2017}), as well as the meaning of effective information in general (e.g., perhaps it is somehow merely capturing "explanatory" causation rather than real causation \cite{dewhurst_causal_2021}). Meanwhile, the integrated information has been criticized for being one of many possible measures \cite{tegmark_improved_2016, mediano_beyond_2019}, and unsubstantiated from its axioms \cite{bayne_axiomatic_2018}. While there are counterarguments to these specific criticisms of info-theoretic accounts of causation, it is a reasonable question whether causal emergence is a general phenomenon or some highly peculiar quirk of these measures and background assumptions, as this would limit its relevancy significantly.

There are already some reasons to think causal emergence is indeed a broader phenomenon. For example, recent evidence has indicated that the synergistic and unique information component of the mutual information can be greater at macroscales (while the redundant information component is lower) \cite{varley_emergence_2021}, and there have been other causal emergence-based approaches to the partial information decomposition as well \cite{mediano_greater_2021, rosas_reconciling_2020}.

Here we provide evidence for widespread generality of causal emergence as a phenomenon. We show that across a dozen popular historical measures of causation from different fields, causal emergence universally holds true under many different conditions and assumptions as to how the measures are applied. That is, instances of emergent macroscale causation can be detected by the majority of independent measures of causation---at least, all of those that we considered. The widespread nature of causal emergence is because most measures of causation are based on a small set of primitives: specifically, sufficiency and necessity, along with their generalizations (which we provide here) of determinism and degeneracy, respectively. All these causal primitives can be improved at a macroscale. Therefore, all the measures also demonstrate causal emergence (indeed, we find that effective information is the most conservative measure of those we analyzed). This is all despite the fact that macroscales are simply dimension-reductions of microscales. So while two scales may both be valid descriptions of a system, one may possess stronger causation (the interpretation of which, whether as more causal work, information, or explanation, depends on the measure of causation itself). Yet causal emergence is not trivially universal either. It is system-dependent: in many cases, specifically those without any uncertainty in microscale system dynamics, causal reduction dominates.

First, in Section \ref{sec:primitives}, we define causal primitives along with the formal language of cause and effect we will use throughout. In Section \ref{sec:causation_measures}, we overview twelve independently-proposed measures of causation (several of which end up being identical, as we show). In Section \ref{sec:measures_are_sensitivity}, we highlight how the behavior of the measures is based on causal primitives using a simple bipartite Markov chain model. In Section \ref{sec:macroscale_causation}, we directly compare macroscales to microscales across all the causal measures using the bipartite model, and find widespread evidence for causal emergence across all the measures.

In the Discussion, we overview how the consilience of causation we've revealed can provide a template for an objective understanding of causation, and discuss the beginnings of the scientific subfield of emergence.

\section{\label{sec:primitives}Formalizing causation and causal primitives}

First, a note on terminology. We must use a general enough one that it can incorporate a number of different notions of causation from different fields. Therefore, we focus on a given a space $\Omega$, i.e., the set of all possible occurrences. In this space, we can consider causes $c\in\Omega$ and effects $e\in\Omega$, where we assume causes $c$ to precede effects $e$, so that we also speak of a set of causes $C \subseteq \Omega$ and of effects $E \subseteq \Omega$.

As we will later be applying these measures in Markov chains, we can consider the space $\Omega$ to be a state-space and $c$ or $e$ as states. The set of causes and effects can be related probabilistically via transition probabilities $P(e \mid c)$, which specifies the probability of obtaining a candidate effect $e$, given that a candidate cause $c$ actually occurred.

As we will see, in order to gauge causation, we will have to evaluate counterfactuals of $c$, and consider the probability of obtaining the effect $e$ given that $c$ didn't occur. We will write this probability $P(e \mid C\backslash c)$, where $C \backslash c$ stand for the complement of $c$, by which we mean the probability of $e$ given that any cause in $C$ could have produced $e$ except for $c$. 
Note that although conventionally written $P(e)$ we will write $P(e \mid C)$ to underscore the following notion: namely, that to meaningfully talk about $P(e \mid C)$ (and $P(e \mid C\backslash c)$), a further distribution over $C$ must be specified. That is:
 
 $$ P(e \mid C) = \sum_{c\in C} P(c) P(e \mid c)$$
 
where there is some assumption of a distribution $P(C)$. This assumption is necessary because, unlike terms like $P(e \mid c)$ which are stated in some transition probability matrix (TPM) or system description, terms like $P(c)$ or $P(e \mid C \backslash c)$ need to be explicitly defined (e.g., what is the distribution of the effects when $c$ \textit{didn't} occur?). In Section \ref{sec:intervention_distributions} we overview how $P(C)$ itself is defined via an intervention distribution, which is necessary to specify for the application of measures of causation, although not their definitions.  Therefore, we simply assume a particular $P(C)$ is defined for the following measures of causation. Note that in examining counterfactual probabilities like $P(e \mid C \backslash c)$ it implies that $P(C)$ is restricted to exclude $c$ and normalized.

%   This requires an idea of what the distribution over $C$ is, which is best operationalized via the idea of a distribution of interventions \cite{hoel_when_2017}. This distribution $P(C)$, which we call the
  
%   Therefore, this is an operationalization of the idea of interventions \cite{pearl_causality_2009, woodward_making_2005}.

%  \emph{intervention distribution}:

%  In the case of examining measures of necessity or degeneracy, which involve some specification of a counterfactual, like $P(E \mid C \backslash c)$, then the same intervention is taken but without that specific $c$ and then normalized such that the sum = 1.

%  together with the transition probabilities $P(e \mid c)$ allows us to compute the counterfactuals in the following manner:
% $$ P(e \mid C \backslash c) = \sum_{c' \in C\backslash c} P(c') P(e \mid c')$$

%  Note that $P(e \mid c)$ is generally easier to derive---given a particular $c$, the $P(e \mid c)$ is the frequency by which $e$ follows $c$. But what is $P(c)$, or $P(C \backslash c)$, in a measure of causation? Indeed, measures of causation often rely on terms like $P(e \mid C \backslash c)$, i.e., the distribution over the effects when $c$ \textit{didn't} occur. Without a specified $P(C)$ this is not defined. An intervention distribution, which specifies a probability distribution over $C$, that is, $P(C)$, is necessary.

\subsection{\label{sec:primitives_formalization}Formalizing sufficiency and necessity}

\begin{figure}[ht]
  \includegraphics[width=1\textwidth]{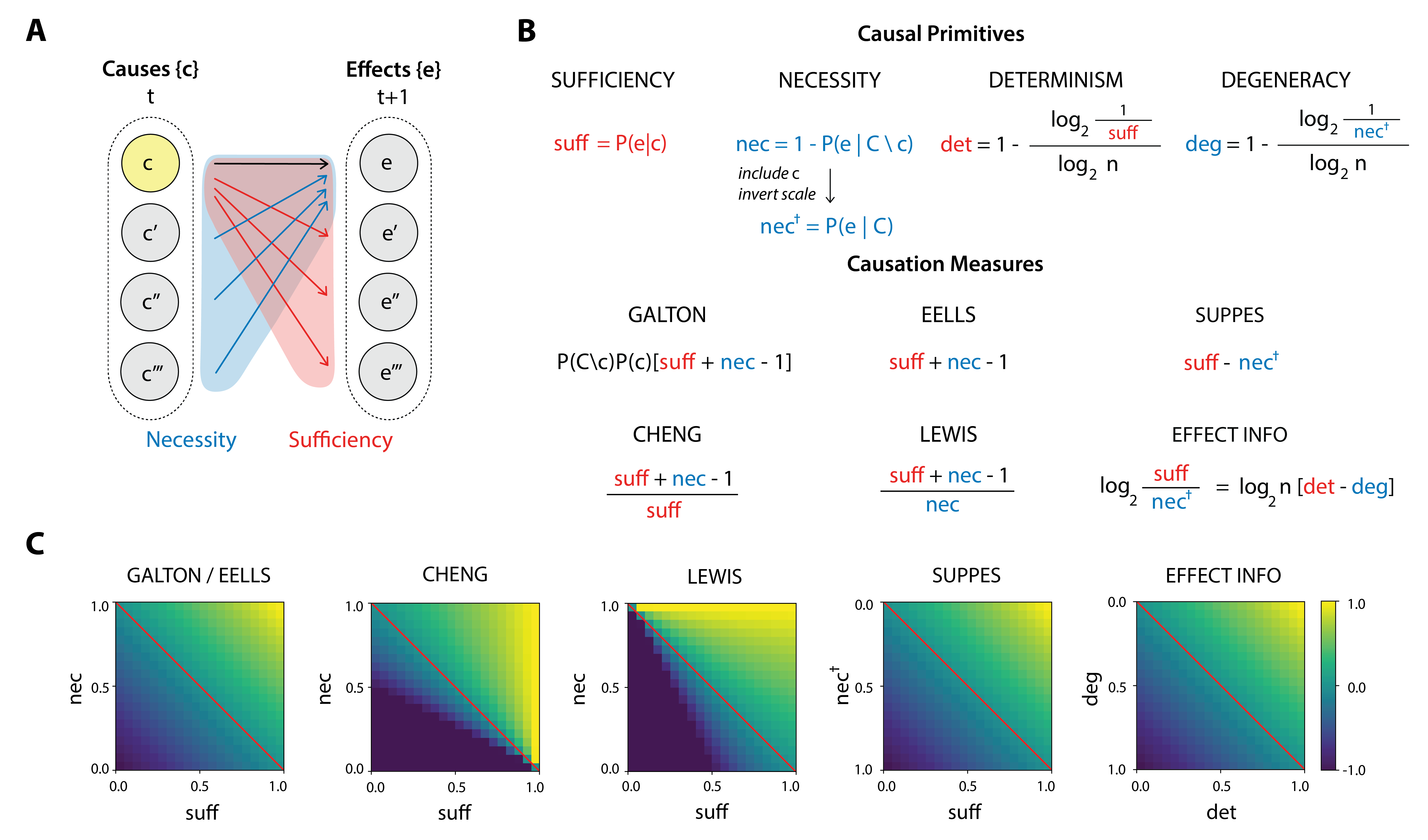}
  \caption{\textbf{Causal primitives and causation measures.} (A) Schematic representation of causation as a relation between occurrences (or events) connecting a set of causes to a set of effects. Each individual candidate cause ($c, c', ...$) and candidate effect ($e, e', ...)$ is depicted in a circle, while their sets $C$ and $E$ are marked as the enclosing dotted line. Causes and effects are assumed to be temporally ordered, with the former preceding the latter, hence are indexed at a time $t$ and a time $t+1$, respectively. Given a pair of a candidate cause $c$ and a candidate effect $e$, the relation between $c$ and $e$ can be analysed in terms of the causal primitives of sufficiency and necessity. On one hand, one can assess whether $c$ is \emph{sufficient} to bring about $e$, or whether $c$ can instead transition to other effects in $E$ (region shaded red); on the other hand, one can ask whether $c$ is \emph{necessary} for $e$ to obtain, or instead whether other causes in $C$ could also produce $e$ (region shaded blue). (B) The functional dependence of the causation measures on the causal primitives is highlighted (sufficiency and determinism in red, necessity and degeneracy in blue). On the top are the formulas of the causal primitives and on the bottom, the formulas of some of the causation measures written in terms of the causal primitives (measures like the Bit-flip and Lewis' closest possible world are not shown because they rely on additional structure, e.g. distances between occurrences). (C) Behavior of the causation measures for different combinations of the causal primitives. Heatmaps show causation as a function of the causal primitives (using $n=2$). For Suppes and effect information the y axis is inverted to highlight the similarity with the other measures.}
  \label{fig:figure_1}
\end{figure}

Causation should be viewed not as an irreducible single relation between a cause and an effect but rather as having two dimensions: that of sufficiency and necessity \cite{pearl_causality_2009, nadathur_causal_2020}.

For any cause $c$, we can always ask, on one hand, how sufficient $c$ is for the production of an effect $e$. A sufficient relation means that whenever $c$ occurs, $e$ also follows (Figure \ref{fig:figure_1}A, red region). Separably, we can also ask how necessary $c$ is to bring about $e$, that is, whether there are different ways then through $c$ to produce $e$ (Figure \ref{fig:figure_1}A, blue region). Yet these properties are orthogonal: a cause $c$ may be sufficient to produce $e$, and yet there may be other ways to produce $e$. Similarly, $c$ may only sometimes produce $e$, but is the only way to do so.

We refer to sufficiency and necessity as \textit{causal primitives}. This is because, as we will show, popular measures of causation generally put these two causal primitives in some sort of relationship (like a difference or a ratio). This ensures such measures are mathematically quite similar, indeed, sometimes unknowingly identical.

First we must define the primitives formally. To start, we associate the sufficiency of the cause $c$ to the probability:

$$ suff(e,c) = P(e \mid  c) $$

which is 1 when $c$ is fully sufficient to produce $e$. This allows for degrees of sufficiency (e.g., a cause might bring about its effect only some of the time), which is important because many measures of causation rely on probability raising or difference making.

Comparably, the necessity of the cause for the effect we associate with the probability:

$$ nec(e,c) = 1 - P(e \mid C \backslash c) $$

which gives "the probability of not-e given the probability of not-c." Necessity is 1 when $c$ is absolutely necessary for $e$. In such cases there is no other candidate cause but $c$ that could produce $e$. Note that some definition of counterfactuals needs to be made explicit for the calculation of necessity, unlike sufficiency (more on this in later sections, where possible counterfactuals are represented as performable interventions).

\subsection{\label{sec:primitives_generalization}Determinism and degeneracy as generalizations of sufficiency and necessity}

The two causal primitives of sufficiency and necessity each have a generalization. These are the determinism and degeneracy coeffients \cite{hoel_quantifying_2013}. Specifically, the determinism coefficient is a generalized notion of sufficiency, while the degeneracy coefficient is a generalized notion of necessity. These generalizations will prove useful in two ways: a) they provide a more general version of the original primitive, and b) some measures of causation are based off of determinism and degeneracy instead of sufficiency and necessity.

We can define the determinism as the opposite of noise (or randomness), that is, the certainty of causal relationships. Specifically, it is based on the entropy of the probability distribution of the effects of a cause:

$$ H(e\mid c) = \sum_{e \in E}P(e \mid c)\log_2\frac{1}{P(e \mid c)}  $$

This entropy term is zero if a cause has a single effect with $P = 1$, and the entropy is maximal, i.e. $\log_{2}n$, if a cause has a totally random effect. We therefore define the determinism of a cause $c$ to be $\log_{2}(n) - H(e\mid c)$. Note that this is different than the mere sufficiency, although is also based on the sufficiency $P(e\mid c)$. To see their difference, let us consider a system of four states $\Omega = \{a, b, c, d\}$,  wherein state $a$ transitions to the other states $b$, $c$, or $d$, and also back to itself, $a$, with probability $1/4$ each. The average sufficiency of $a$'s transitions would be $1/4$. However, the determinism of $a$ would be zero, since there is no difference between $a$ and randomly generating the next state of the system.

Unlike sufficiency, the determinism is a property of a cause, not a particular transition (although the contribution of each transition to the determinism term can be calculated). And unlike sufficiency, the determinism term is influenced by the number of considered possibilities. Generally, we normalize the term to create a determinism coefficient that ranges, like the sufficiency, between 0 (fully random) and 1 (fully deterministic), for a given cause:

$$ det(c) = 1 - \frac{H(e \mid c)}{\log_2 n}$$

% $$ det(c) = \sum_{e\in\Omega} P(e | c) det(e, c) = 1 - \frac{H(e \mid c)}{\log_2 n}$$.

And with this in hand, we can define a determinism coefficient for individual transitions as:

$$det(e, c) = 1 - \frac{\log_2 \frac{1}{P(e \mid c)}}{\log_2 n}$$

% We can further define a bit transition-based determinism value as:
% $$ \log n - \log\frac{1}{P(e \mid c)} $$
% which is 0 if $c$ can produce $e$ with the same power as a maximum entropy distribution where $P(e \mid c)=1/n$.

% And by averaging across all possible effects weighed by their transition probabilities we get the determinism coefficient associated to a particular state or cause:
% $$ det(c) = \sum_{e\in\Omega} P(e | c) det(e, c) = 1 - \frac{H(e \mid c)}{\log_2 n}$$

% By doing the joint average we recover the original determinism coefficient proposed in \cite{hoel_quantifying_2013} (though here defined inversely):
% $$ det = \sum_{c,e\in\Omega} P(e, c) det(e, c) = 1 - \frac{\langle H(e \mid c)\rangle}{\log_2 n}$$.
as well as a system-level determinism coefficient:

$$ det = \sum_{c\in C} P(c) \ det(c) = \sum_{e\in E, \ c\in C} P(e,c) \ det(e, c) = 1 - \frac{\sum_{c\in C} P(c) \ H(e \mid c)}{\log_2 n} $$

% To generalize the necessity, which gauges the counterfactuals of a certain transition from a candidate cause c to a candidate effect e, we can simply expand the counterfactual space to include the original c as well. The heart of the degeneracy is therefore that it is based on P(e | C) rather than 1 − P(e | C \ c), a sort of "anti-necessity". we can formalize the degeneracy using another entropy term, this time based on P(e | C) instead of P(e | c) :

Degeneracy is the generalization of necessity. It is zero when no effect has a greater probability than any other (assuming an equal probability across the full set of causes). Degeneracy is high if certain effects are "favored" in that more causes lead to them (and therefore those causes are less necessary). It is also based on an entropy term:

$$H(e\mid C) = \sum_{e \in E}P(e \mid C)\log_2\frac{1}{P(e \mid C)} $$

and the degeneracy coefficient of an individual effect is given by:

$$ deg(e) = 1 - \frac{\log_2 \frac{1}{P(e\mid C)}}{\log_2 n} $$

while the system-level degeneracy coefficient is:

$$ deg = \sum_{e\in E} P(e \mid c) \ deg(e) = 1 - \frac{H(e \mid C)}{\log_2 n} $$

\section{\label{sec:causation_measures}Measures of causation are based on causal primitives}

In the following section, we demonstrate how  the basic causal primitives of sufficiency and necessity or their generalized forms of determinism and necessity underlie the independent popular measures of causation we examined.

\subsection{\label{sec:constant_conjunction} Humean constant conjunction}

One of the earliest and most influential approaches to a modern view of causation was David Hume's regularity account. Hume famously defined a cause as "an object, followed by another, and where all the objects, similar to the ﬁrst, are followed by objects similar to the second" \cite{hume_enquiry_1748}. In other words, causation stems from patterns of succession between events \cite{illari_causality_2014}. 

Overall, the "constant conjunction" of an event $c$ followed by an event $e$, would lead us to \textit{expect} $e$ once observing $c$, and therefore infer $c$ to be the cause of $e$. There are a number of modern formalisms of this idea. Here we follow Judea Pearl, who interprets Hume's notion of "regularity of succession" as amounting to what we today call correlation between events \cite{pearl_causality_2009}. This can can be formalized as the observed statistical covariance between a candidate cause $c$ and effect $e$:

$$ Cov(X, Y) = E(X Y) - E(X)E(Y) $$

If we substitute the indicator function $X_c$ (and $Y_e$), which is 1 if $c$ (respectively $e$) occurs and 0 otherwise, in the equation above we obtain:

\begin{align*}
    Cov(X_c, Y_e) & = P(c, e) - P(c)P(e) \\
    & = P(c)P(e \mid c) - P(c)[P(c)P(e \mid c) + P(\bar{c})P(e \mid C \backslash c)] \\
    & = P(e \mid c) P(c)[1 - P(c)] + P(c)P(C \backslash c)P(e \mid C \backslash c) \\
    & = P(e \mid c) P(c)P(C \backslash c)] + P(c)P(C \backslash c)P(e \mid C \backslash c) \\
    & = P(c)P(C \backslash c)[P(e \mid  c) - P(e \mid  C \backslash c)])
\end{align*}
 
Where we used the fact that $P(e \mid C)$) can be decomposed into two weighted sums, i.e. over $c$ and over $C \backslash c$. Following other's nomenclature \cite{fitelson_probabilistic_2010} we call this the "Galton measure" of causal strength, since it closely resembles the formalism for heredity of traits in biology, and also is a form of the statistical co-variance:

\begin{align*}
    CS_{Galton}(e, c) & = P(c)P(C \backslash c)[P(e \mid  c) - P(e \mid  C \backslash c)] = P(c)P(C \backslash c)[suff(e, c) + nec(e, c) - 1]
\end{align*}

It's worth noting that such a regularity account can be stated in ways that involve causal primitives, as can be seen above.

\subsection{\label{sec:eells}Eells's measure of causation as probability raising}

Ellery Eells proposed that a condition for $c$ to be a cause of $e$ is that the probability of $e$ in the presence of $c$ must be higher than its probability in its absence: $P(e \mid  c) > P(e \mid  C \backslash c)$ \cite{eells_probabilistic_1991}. This can be formalized in a measure of causal strength as the difference between the two quantities:

$$ CS_{Eells} = P(e \mid  c) - P(e \mid  C \backslash c) = suff(e, c) + nec(e, c) - 1$$

When $CS_{Eells} < 0$ the cause is traditionally said to be a negative or preventive cause \cite{illari_causality_2014}, or in another interpretation, such negative values should not be considered a cause at all \cite{albantakis_what_2019}.

\subsection{\label{sec:suppes}Suppes's measure of causation as probability raising}

Another notion of causation as probability raising was defined by Patrick Suppes, a philosopher and scientist \cite{suppes_probabilistic_1968}. Translated into our formalism, his measure is:
  
$$ CS_{Suppes}(c, e) = P(e \mid  c) - P(e\mid C) = suff(e, c) - nec^{\dagger}(e)$$

The difference between the $CS_{Eells}$ and $CS_{Suppes}$ measures involves a shift from measuring how causally \textit{necessary} $c$ is for $e$---whether it can be produced by other causes than $c$---to assessing how \textit{degenerate} is the space of ways to bring $e$ about. Both are valid measures, and in fact turn out to be equivalent in some contexts \cite{hitchcock_probabilistic_2018}.

Note that we can extend the conditional probability $P(e \mid C \backslash c)$ to $P(e \mid C)$, including $c$ itself. If so, we are considering whether $e$ can be produced not just in the absence of $c$, but all the ways, including via $c$ itself, that $e$ can occur. Therefore, another version can be defined as:

$$ CS_{Suppes_{II}}(c, e) = \frac{P(e \mid  c)}{P(e \mid C)}$$

\subsection{\label{sec:cheng}Cheng's causal attribution}

Patricia Cheng has proposed a popular psychological model of causal attribution, where reasoners go beyond assessing pure covariation between events to estimate the "causal power" of a candidate cause producing (or preventing) an effect \cite{cheng_causes_1991}. In this account, the causal power of $c$ to produce $e$ is given by:

$$ CS_{Cheng}(c, e) = \frac{P(e\mid c) - P(e\mid C \backslash c)}{1 - P(e \mid  C \backslash c)} = \frac{suff(e, c) + nec(e, c) - 1}{nec(e, c)}$$

Cheng writes: "The goal of these explanations of $P(e\mid c)$ and $P(e\mid C \backslash c)$ is to yield an estimate of the (generative or preventive) power of $c$...." While originally proposed as a way to estimate causes from data based off of observables, it's worth noting that, in our application of this measure, we have access to the real probabilities given by the transition probability matrix $P(e\mid c)$, and the measure therefore yields a true assessment of causal strength, not an estimation.

\subsection{\label{sec:lewis}Lewis's counterfactual theory of causation}

Another substantive and influential account of causation based on counterfactuals was given by philosopher David Lewis \cite{lewis_causation_1973}. Lewis defines a cause as if given events $c$ and $e$ took place, $c$ can be said to be a cause of $e$ if it is the case that if $c$ hadn't occurred, then $e$ would not have occurred. Lewis also extended his theory for "chancy worlds", where $e$ can follow from $c$ probabilistically \cite{lewis_postscripts_1986}. 

Following \cite{fitelson_probabilistic_2010} we formalize Lewis's causal strength as the ratio:

$$ \frac{P(e \mid  c)}{P(e \mid  C \backslash c)} $$

This definition is also known as "relative risk:" "it is the risk of experiencing $e$ in the presence of $c$, relative to the risk of $e$ in the absence of $c$ " \cite{fitelson_probabilistic_2010}. This measure can be normalized to obtain a measure ranging from -1 to 1 using the mapping $p/q \rightarrow  (p - q)/p$ as:

$$ CS_{Lewis}(c, e) = \frac{P(e \mid  c) - P(e \mid  C \backslash c)}{P(e \mid  c)} = \frac{suff(e, c) + nec(e, c) - 1}{suff(e, c)}$$

Again we see that Lewis's basic notion, once properly formalized, is based on the comparison of a small set of causal primitives. Note that this definition doesn't rely on a specification of a particular possible world. In other work, Lewis specifies that the counterfactual not-$c$ is taken to be the closest possible world where $c$ didn't occur. That notion, which specifies a rationale for how to calculate the counterfactual, is formalized in Section \ref{sec:Lewis_closest}. 

\subsection{\label{sec:Judea_Pearl}Judea Pearl's measures of causation}

If our claim for consilience in the study of causation is true, then authors should regularly rediscover previous measures. Indeed, this is precisely what occurs. Consider Judea Pearl, who in his work on causation has defined the previous measures $CS_{Eells}$, $CS_{Lewis}$, and $CS_{Cheng}$ (in some of these terms apparently knowingly, in others not).

Within his structural model semantics framework \cite{pearl_causality_2009}, he defines the "probability of necessity" as the counterfactual probability that $e$ would not have occurred in the absence of $c$, given that $c$ and $e$ did in fact occur, which in his notation is written as $\textrm{PN} = P(\bar{e}_{\bar{c}} \mid  c, e)$ (where the bar stands for the complement operator, i.e. $\bar{c} = C \backslash c$). Meanwhile, he defines the "probability of sufficiency" as the capacity of $c$ to produce $e$ and is defined as the probability that $e$ would have occurred in the presence of $c$, given that $c$ and $e$ didn't occur: $ \text{PS} = P(e_c \mid  \bar{c}, \bar{e})$.

Finally, both aspects are combined to measure both the sufficiency and the necessity of $c$ to produce $e$ as $\textrm{PNS} = P(e_c, \bar{e}_{\bar{c}})$, such that the following relation holds: $\textrm{PNS} =  P(e, c)\textrm{PN} + P(\bar{e}, \bar{c})\textrm{PS}$.

% The three measures should satisfy:

% Under exogeneity ($P(e\mid  do(c)$) is an intervention distribution; ${E_c, E_{\bar{c}}} \perp X $) we have:

% $$ \textrm{PN} = \frac{\textrm{PNS}}{P(e \mid  c)} $$

% $$ \textrm{PS} = \frac{\textrm{PNS}}{P(\bar{e}\mid  \bar{c})} $$

Under conditions of exogeneity of $c$ relative to $e$ (which renders Pearl's counterfactual $P(e_c)$, i.e. the causal effect of $c$ on $e$, computable from $P(e \mid c$)) and monotonicity of $e$ relative to $c$ (which roughly means $c$ does not prevent $e$ from happening), the measures are given by:

$$ \textrm{PNS} = P(e \mid  c) - P(e \mid  C \backslash c) $$

$$ \textrm{PN} = \frac{P(e \mid  c) - P(e \mid  C \backslash c)}{P(e \mid  c)}$$

$$ \textrm{PS} = \frac{P(e \mid  c) - P(e \mid  C \backslash c)}{1 - P(e \mid  C \backslash c)} $$

where we can recognize that PNS $= CS_{Eells}$, PN $=CS_{Lewis}$ and PS $=CS_{Cheng}$ \cite{fitelson_probabilistic_2010}. That is, Pearl independently recreated previous measures.

Yet we find Pearl's terminology confusing. Therefore, we reserve terms like "probability of sufficiency" to mean the original sufficiency, $P(e \mid c)$, and likewise for the probability of necessity $1 - P(e \mid C \backslash c)$. We also continue to refer to the measures by the names of their respective original authors to further distinguish them and preserve origin credit.

Overall this consilience should increase our confidence that measures based on the combinations of causal primitives are good candidates for assessing causation.

\subsection{\label{sec:Lewis_closest}Closest possible world causation}

As state previously, David Lewis traditionally gives a counterfactual theory of causation wherein the counterfactual is specified as the closest possible world where $c$ didn't occur \cite{lewis_causation_1973}. In order to formalize this idea, we need to add further structure beyond solely probability transitions. That is, such a measurement requires a notion of distance between possible states of affairs (or "worlds"). One simple way to do is use binary state labels of states to induce a metric using the Hamming distance \cite{floridi_information_2010}, which is the number of bit flips needed to change one binary string into the other. In this way we induce a metric in a state-space so that we can define Lewis notion of a closest possible world:

$$ D_H(x, y) = \sum\limits_{i}^{N} \lvert x_i - y_i \rvert $$

where $x$ and $y$ are two state labels with $N$ binary digits (e.g. $x=0001$ and $y=0010$, $N=4$, such that $D_{H}(x, y) = 2$). With such a distance notion specified the counterfactual taken as the "closest possible world" where $c$ didn't occur is given by:

$$ \bar{c}_{CPW} = \min\limits_{c'} D_H(c, c')$$

And with this in hand, we can define another measure based closely on Lewis's account of causation as reasoned about from a counterfactual of the closest possible world:

$$ CS_{Lewis \ CPW} = \frac{P(e \mid c) - P(e \mid \bar{c}_{CPW})}{P(e \mid c)} $$

\subsection{\label{sec:bit_flip_measures}Bit-flip measures}

Another measure that relies on a notion of distance between states is the idea of measuring the amount of difference created by a minimal change in the system. For instance, the outcome of flipping of a bit from some local perturbation. In \cite{daniels_criticality_2018} such a measure is given as "the average Hamming distance between the perturbed and unperturbed state at time $t + 1$ when a random bit is flipped at time $t$". While originally introduced with an assumption of determinism, here we extend their measure to non-deterministic systems as:

$$ CS_{bit-flip}(e, c) = \frac{1}{N}\sum_{i}^N\sum_{e' \in E}P(e'\mid c_{[i]})D_H(e, e')$$

where $c_{[i]}$ correspond to the state where the $i^{th}$ bit is flipped (e.g., if $c=000$, then $c_{[3]}=001$).

% We will develop this idea present in Lewis notion of closest possible world and Walker's bit flip measure as a way of exploring the counterfactual space $\bar{c}$ in the next section where the choice of an input distribution $P(c)$ will come into play.

\subsection{\label{sec:actual_causation}Actual causation and the effect information}

Recently a framework was put forward \cite{albantakis_what_2019} for assessing actual causation on dynamical causal networks, using information theory. According to this framework, a candidate cause must raise the probability of its effect compared to its probability when the cause is not specified (again, we see similarities to previous measures). The central quantity is the \textit{effect information}, given by:

$$ ei(c, e) = \log_2\frac{P(e \mid c)}{P(e\mid C)} = \log_2 n [det(e,c) - deg(c)]$$

Note that the effect information is actually just the log of $CS_{Suppes_{II}}$, again indicating consilience as measures of causation are re-discovered by later authors. It is also the individual transition contribution of the previously defined "effectiveness" given in \cite{hoel_quantifying_2013}.

% Therefore, in order to relate it to other measures, we give an equivalent definition of an effectiveness coefficient:

% $$ eff(e, c) = deg(e) - det(e, c) = \frac{1}{\log_2 n} \left(\log_2\frac{1}{P(e \mid c)} - \log_2\frac{1}{P(e \mid c)}\right) = \frac{1}{\log_2 n} \log_2\frac{P(e \mid c)}{P(e \mid c)} = \frac{1}{\log_2 n} \rho(c, e) $$

The effect information is thus on one hand a bit-measure version of the probabilistic Suppes measure, and on the other an non-normalized difference between degeneracy and determinism.

\subsection{\label{sec:effective_information}Effective information}

The effective information ($EI$) was first introduced by Giulio Tononi and Olaf Sporns as a measure of causal interaction, in which random perturbations of the system are used in order to go beyond statistical dependence \cite{tononi_measuring_2003}. It was rediscovered without reference to prior usage and called "causal specificity" \cite{griffiths_measuring_2015}.

% Based on the previous section, a state-dependant effect information $ei(c)$ can be introduced as a weighted average of the effect information of a cause $c$ across all its potential effects. Alternatively, $ei(c)$ can be formulated, as in \cite{hoel_quantifying_2013}, as the Kullback-Leibler divergence between so-called constrained effect repertoire ($P(e \mid c)$) and the unconstrained effect repertoire ($P(e \mid C$). The equivalence of both definitions can be readily seen:

% \begin{align*}
%     ei(c) & = D_{KL}(P(e \mid c) \ \| \ P(e \mid c)) \\
%         & =  \sum_{e \in E}P(e\mid c)\log_2\frac{P(e\mid c)}{P(e\mid C)} \\
%         & = \sum_{e \in E}P(e\mid c) \ ei(c, e)
% \end{align*}

% Where the divergence between two distributions is given by $D_{KL}(P \| Q) = \sum_x P(x) \log_2\frac{P(x)}{Q(x)}$. The effect information $ei(c)$ thus measures how much information each occurrence $c$ specifies about its (positive and negative) effects, above and beyond the average effect.
The effective information is simply the expected value of the effect information over all the possible cause-effect relationships of the system:

$$ EI = \sum_{e \in E, c \in C} P(e,c) ei(c, e) = \log_2 n [det - deg]$$

As a measure of causation, $EI$ captures how effectively (deterministically and uniquely) causes produce effects in the system, and how selectively causes can be identiﬁed from effects \cite{hoel_quantifying_2013}.

Effective information is an assessment of the causal power of $c$ to produce $e$ – as measured by the \textit{effect information} – for all transitions between possible causes and possible effects, considering a maximum-entropy intervention distribution on causes (the notion of an intervention distribution is discussed in Section 3.4). More simply, it is the non-normalized difference between the system's determinism and degeneracy. Indeed, we can normalize the effective information by its maximum value, $\log_2 n$, to get the \textit{effectiveness} of the system:

$$ eff = det - deg = \frac{EI}{\log_2 n} $$

\subsection{\label{sec:Behavior}Summary}

Across every measure of causation we examined the two primitives (sufficiency and necessity), or alternatively their generalized forms (determinism and degeneracy), are explicitly put in some relationship, often that of a difference or ratio or trade-off (Figure \ref{fig:figure_1}, panels B and C). The only measure that lacked an explicitly obvious basis in causal primitives was the bit-flip measure, but as a measure of sensitivity to perturbation it seems likely there is some basis or relationship (we did not seek out a decomposition).

We are not the first to point out that causation has two dimensions: for instance, Judea Pearl \cite{pearl_causality_2009} states: "Clearly, some balance must be struck between the necessary and the sufficient components of causal explanation." Also J. L. Mackie, although not proposing a quantitative measure of causal strength, famously considers both a necessity and a sufficiency aspect in his proposal of a INUS condition that causes should satisfy, namely being an (i)nsufficient but (n)ecessary part of a condition which is itself (u)nnecessary but (s)ufficient for an effect to occur \cite{mackie_causes_1965}. However, to our knowledge this is the first time a full set of popular measures has been assessed in this light, and so we state it explicitly: substantial consilience in measures of causation indicates we should expect measures of causal strength to be based on \textit{both} causal primitives.

\section{\label{sec:measures_are_sensitivity}Measures of causation are sensitive to causal primitives}

\subsection{\label{sec:model_system}Model system}

\begin{figure}[ht]
  \includegraphics[width=1\textwidth]{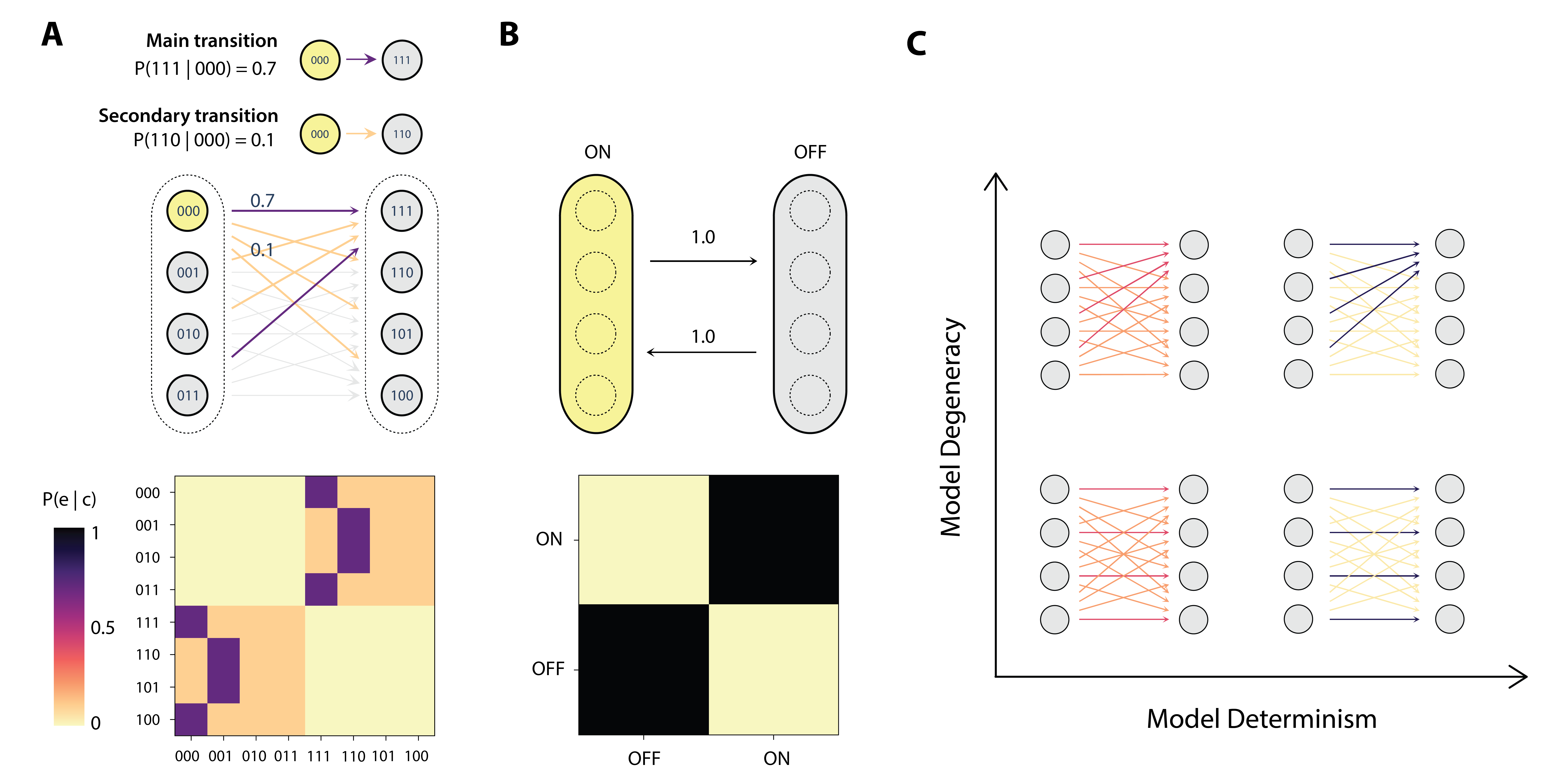}
  \caption{\textbf{A simple bipartite Markov chain model for studying causal measures.} (A) Microscale model of the bipartite Markov chain with 8 microstates where microstates transition back-and-forth between two groupings (left and right). On the top, a representation of the state-space with binary labels is shown, with the dotted line indicating the natural macrostate. The possible microstate transitions from the group in the left to the group in the right are represented by the arrows (the transitions from right to left are omitted). A main transition from $000$ to $111$ is highlighted, and contrasted with a secondary transition from $000$ to $110$, which intuitively has a lower causal strength. The relevant transitions for evaluating the causal primitives (suffiency and necessity) for the main transition are color coded by the probability according to the probability transition matrix (TPM) shown in the bottom. (B) Macro model of bipartite Markov chain obtained by coarse-graining the original microstates into the two groupings ($\textrm{ON} = \{000, 001, 010, 011\}$ and $\textrm{OFF} = \{111, 110, 101, 100\}$). The obtained macro transition probability matrix (TPM) describes the deterministic transition between the OFF and ON states ($P(\textrm{ON}_{t+1} \mid \textrm{OFF}_t)=1$ and $P(\textrm{OFF}_{t+1} \mid \textrm{ON}_{t})=1$) and is independent of the model parameters that control the determinism and degeneracy of the microscale.  (C) Graphical representation of the bipartite model state-space (as shown in panel A) for different values of the model's microscale degeneracy and determinism.}
    \label{fig:figure_2}

\end{figure}
    
In order to examine the behavior of measures of causation presented in the previous section, we make use of a simple model. It was chosen because it allows us to parametrically vary the causal primitives of determinism ($det$) and degeneracy ($deg$) in order to see how the measures of causation change under uncertainty. We make use of a simple bipartite Markov chain model where the microstates of the system  oscillate back and forth between two groups. What is important to keep in mind is that this model is a) bipartite, and b) that we can vary these bipartite connections to either increase the determinism (increasing the average probability of state transition closer to $p=1$) or the degeneracy (increasing the overlap of state transitions, such that transitions cluster in their targets). This allows us to apply the measures of causation under different amounts of uncertainty and different types of uncertainty (like indeterminism vs. degeneracy) and later to also examine causal emergence in such regimes as well. A detailed description of the bipartite model, as well as how we vary these parameters can be found in the Supplementary Information Section \ref{sec:SI_model}. See Figure \ref{fig:figure_2} for a visual representation of the system state-space and transition probability matrix (panel A) and of the different regimes of model architecture we examine (panel C). The code used for calculating the measures of causation as well as assessing causal emergence on the bipartite Markov chain model is available at \url{https://github.com/renzocom/causal_emergence}.

\subsection{\label{sec:intervention_distributions}Applying measures of causation requires defining an intervention distribution}

As we have seen, measures of causation, which can be interpreted as "strength" or "influence" or "informativeness" or "power" or "work (depending on the measure) are based on a combination of causal primitives. However, both the calculations of measures themselves, as well as the causal primitives, involve further background assumptions in order to apply them.

To give a classic example: you go away and ask a friend to water your plant. They don't, and the plant dies. Counterfactually, if your friend had intervened to water the plant, it'd still be alive, and therefore your friend not watering the plant caused its death. However, if the Queen of England had intervened to water the plant, it'd also still be alive, and therefore it appears your plant's death was caused just as much by the Queen of England. This intuitively seems wrong. How do we appropriately evaluate the space of \textit{sensible} counterfactuals or states over which we assess causation? As we will discuss, there are several options. 

Previous research has introduced a formalism capable of dealing with this issue in the form of an \textit{intervention distribution} \cite{hoel_when_2017}. An intervention distribution is a probability distribution over possible interventions that a modeler or experimenter considers. Effectively, rather than considering a single $\textit{do}(\textit{x})$ operator \cite{pearl_causality_2009}, it is a probability distribution over some applied set of them. The intervention distribution fixes $P(C)$, the probability of causes, which is in fact necessary to calculate all the proposed causal measures.

% As another example, the unconditioned effect distribution $P(e\mid C)$ can be thought of as an weighted average of the transition from different states $s$ to $e$ under a particular choice of intervention distribution:

% $$ P(e\mid C) = \sum_{s\in \Omega} P(e\mid s)P(c) = \langle P(e\mid C)\rangle_{\sim C} $$

% Likewise, the counterfactual distribution $P(e\mid C \backslash c)$ can be thought as an weighted average of the transition from different states $s$ to $e$, but restricting or renormalizing the intervention distribution such that $P(c)=0$ so that $c$ is excluded:

% $$ P(e\mid C \backslash c) = \sum_{s\neq c, s\in \Omega} P(e\mid s)P(i) = \langle P(e\mid C)\rangle_{\sim C\backslash c}$$

We point out that there are essentially three choices that a modeler/experimenter has for intervention distributions. The first obvious choice is the \textit{observational distribution}. Also sometimes called an "observed distribution," in the  dynamical systems we're discussing this corresponds to the stationary distribution of states:

$$ P_{obs}(c) = \lim_{n\rightarrow\infty} P^n(e \mid c) $$

In this choice, $P(C)$ is simply based on the system's dynamics itself. However, this choice suffers from serious problems---indeed, much has been made of the fact that analyzing causation must explicitly be about what \textit{didn't} happen, i.e., departures from dynamics, and the observational distribution misses this \cite{pearl_book_2017}. In this case, your plant is dead because your friend didn't water it but you can't even consider what would have happened if they had, since it's not in the observational distribution. In another example: a light-switch would have varying causal power over a light-bulb based entirely on the probability of the person in its house switching it on and off. Another example: a dynamical system with point attractors has no causation under this assumption. This is because the gain from mere observation to perturbing or intervening is lost when the intervention distribution equals the observational distribution. Finally, it is worth noting that definable stationary distributions rarely exist in the real world.

To remedy this, measures of causation often implicitly assume the second choice: an unbiased distribution of causes over $\Omega$, totally separate from the dynamics of the system. In its simplest form, this is described as a maximum-entropy intervention distribution:

$$ P_{maxent}(c) = \frac{1}{n}$$

where $|\Omega| = n$. The maximum-entropy distribution has been made explicit in the calculation of, for instance, Integrated Information Theory \cite{oizumi_phenomenology_2014} or the previously-described effective information of Section \ref{sec:effective_information} \cite{tononi_measuring_2003}. There are a number of advantages to this choice, at least when compared to the observational distribution. First, it allows for the appropriate analysis of counterfactuals. Second, it is equivalent to randomization or noise injection, which severs common causes. Third, it is the maximally-informative set of interventions (in that maximum-entropy has been "injected" into the system).

However, it also has some disadvantages. Using a maximum-entropy intervention distribution faces the difficulty that if $\Omega$ is too large, it might be too computationally expensive to compute. More fundamentally, using $P_{maxent}(c)$ can lead to absurdity (e.g., it assumes that the counterfactual wherein the Queen of England watered the plant is just as equally likely as your friend watering it, thus leading to the paradox wherein your friend is not a necessary cause of your plant's death). That is, $P_{maxent}(c)$, taken literally, involves very distant and unlikely possible states of affairs. However, in cases where the causal model has already been implicitly winnowed to be over events that are considered likely, related, or sensible---such an already constructed or bounded causal model, like a set of logic gates, gene regulations, or neuronal connections---it allows for a clear application and comparison of measures of causation.

We point out there is a third possible construction of an intervention distribution. This is to take a local sampling of the possible world space (wherein locality is distance in possible worlds, states of affairs, the state-space of the system, or even based on some outside non-causal information about the system). There are a number of measures of causation that are constructed around local interventions; one of the earliest and most influential is David Lewis's idea of using the closest possible world as the counterfactual by which to reason about causation \ref{sec:Lewis_closest}. Other examples that implicitly take a local intervention approach includes the bit-flip measure \cite{daniels_criticality_2018} of Section \ref{sec:bit_flip_measures}, as well as the "causal geometry" extension of effective information in continuous systems \cite{chvykov_causal_2020}. We formalize the assumptions behind these approaches as a local intervention distribution, which are possible states of affairs that are similar (or "close") to the current state or dynamics of the system. 

For example, to calculate Lewis's measure, we can compute locality using the Hamming distance \cite{floridi_information_2010}. Rather than simply picking a single possible counterfactual  $\bar{c} \in \Omega$ (which in Lewis's measure would be the closest possible world from Section \ref{sec:Lewis_closest}) we can instead create a local intervention distribution which is a local sampling of states of affairs where $c$ didn't occur. This is equivalent to considering all states which are a Hamming distance less or equal to $\Delta$ from the actual state:

$$
P_{local(c^*)}(c) = \begin{cases}
        \frac{1}{n_{\Delta}}, &\text{if $c \in \Theta_{c^*}$}\\
        $0$ &\text{otherwise}\\
        \end{cases}
$$ 

$$
\Theta_{c^*} = \{s \in \Omega \mid D_H(s, c) \leq \Delta \}
$$

where $n_{\Delta} = |\Theta_{c^*}|$. For example, if we want to locally intervene within a distance $\Delta=1$ around an actual state $c^*=001$, then $\Theta_{c^*} = \{001, 101, 011, 000\}$ and $n_\Delta=4$, such that the intervention distribution is $1/4$ over the four states and $0$ elsewhere.

We note that local interventions avoid many of the challenging edge cases of measuring causation. Therefore, we we use local interventions for our main text figures to highlight their advantages. However, in order to make our points about the consilience of measures of causation, as well as causal emergence, we take an exhaustive approach and consider all three choices of intervention distributions for the dozen measures. It should be stressed that a) the measures again behave quite similarly, even across different choices of intervention distributions, and also b) instances of causal emergence are, as we will show, generally unaffected by choice of intervention distribution.

\subsection{\label{sec:measures_applied}All measures of causation are sensitive to noise}

\begin{figure}[h!]
\begin{center}
    \includegraphics[width=0.70\textwidth]{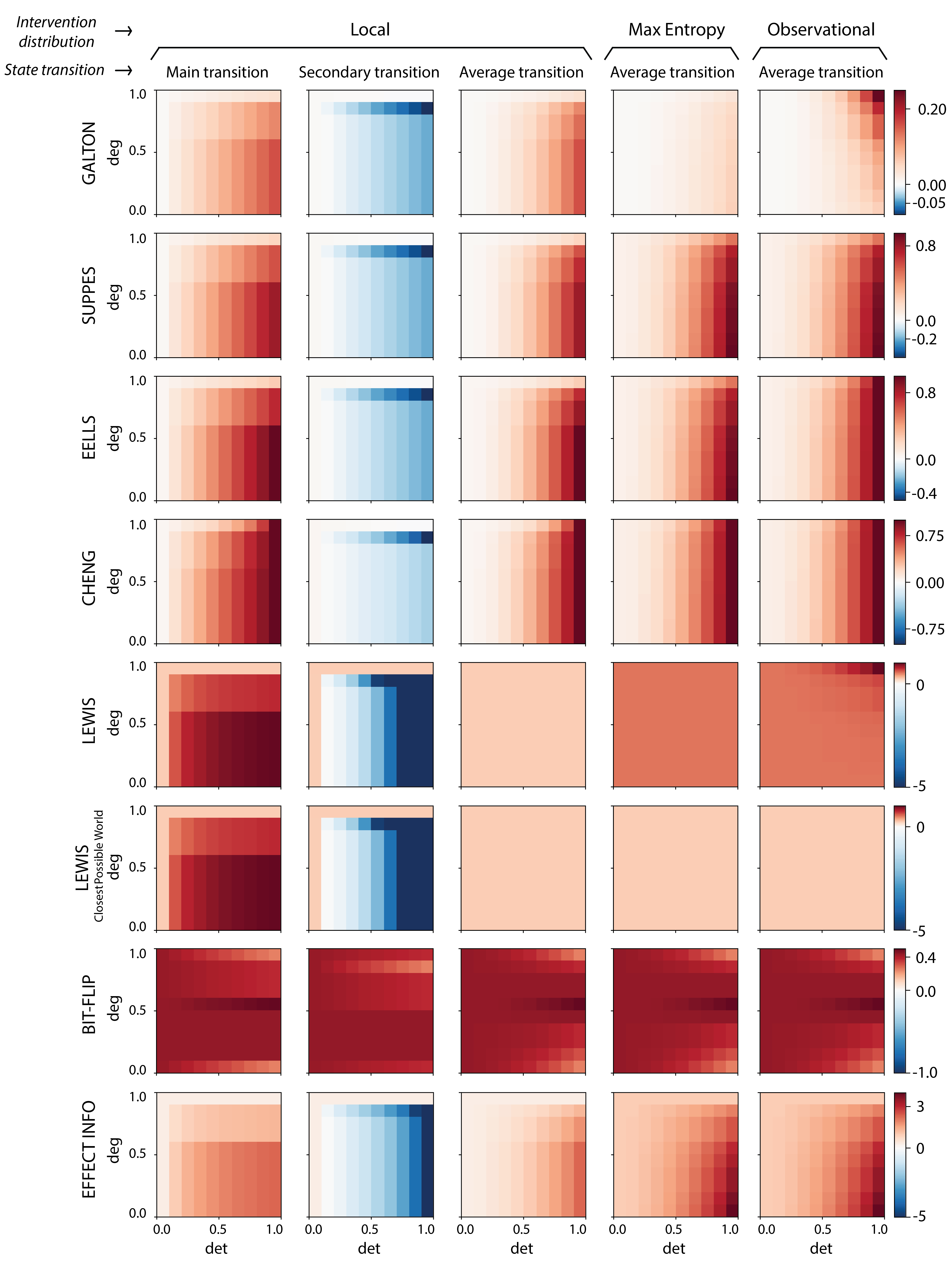}
  \caption{\textbf{Behavior of the causation measures in the model system.} Heatmaps of causal strength are shown for all measures (rows) calculated for the microscale of the bipartite Markov chain model with $n=16$ microstates, 8 states in each macro group ($\Omega_A=\{0000, 0001, 0010, 0011, 0100, 0101, 0110, 0111\}$ and $\Omega_B=\{1111, 1110, 1101, 1100, 1011, 1010, 1001, 1000\}$), at different values of the determinism and degeneracy parameters (see Section \ref{sec:SI_model} in the Supplementary Information for a detailed description). Positive values indicate presence of causal strength and are depicted in red, while negative values correspond to what is known as preemptive or negative causation and are shown in blue. Each measure was calculated for different transitions in the bipartite model: a main transition where a strong causal link is thought to be present ($0000 \rightarrow 1111$), a secondary transition where the causal relationship is supposedly weak ($0000 \rightarrow 1110$), and the average across all state transitions. This average is computed as the joint expectation of the measure $CS(c,e)$ across all transitions using $P(c, e)$ calculated using the transition probability matrix (TPM) and the observational distribution to reflect the expectation of causal strength. The measures were computed using different intervention distributions to assess the counterfactuals: the maximum entropy distribution (all states are uniformly sampled), the stationary distribution (states are sampled according to the observed distribution of the dynamics of the model) and the local distribution (the candidate cause is locally perturbed, so that "close" counterfactuals are sampled). For each measure (row), a common scale is used (shown in the colorbar). The full combinations of intervention distributions and transitions can be found in Supplementary Figure \ref{fig:figure_S2}.} 
\label{fig:figure_3}

\end{center}
\end{figure}

To demonstrate the consilience between measures of causation, as well as their underlying causal primitives, we study their behavior in the model described in section \ref{sec:model_system} under different parameterizations of noise in the form of indeterminism and degeneracy. Due to how we paramaterize determinism and degeneracy, we can simplify looking at every single transition in the model into just two. This is because any given state has a \textit{main transition}, which is the transition of highest probability (e.g., $000 \rightarrow 111$ in Figure \ref{fig:figure_2}A) and its set of \textit{secondary transitions} which are the lower probabilities of transitions (e.g., $001 \rightarrow 111$ in Figure \ref{fig:figure_2}A). When the probability of main transitions equals that of secondary transitions, the system is maximally indeterminate, since all state transitions are a random choice (maximum noise of prediction). This is what is occurring along the $det$ (determinism) axis in Figure \ref{fig:figure_3}. When main effects are stacked on top of a given target, this is increasing the $deg$ (degeneracy axis) (maximum noise in retrodiction). The precise nature of this parameterization and how it reflects the determinism and degeneracy is discussed in Supplementary Section \ref{sec:SI}.

We apply the measures of causation in Section 2 in both a state-dependent and a state-independent manner, since both are common throughout the literature on causation \cite{balduzzi_integrated_2008, halpern_actual_2016, albantakis_what_2019, juel_when_2019, adami_use_2012, timme_tutorial_2018}. We examined the behavior of the measures on specific transitions (such as identifying strong or weak causes) but also their expectation averaged across all transitions, thus covering both individual and global causal properties of the system.

% \begin{align*}
% CS_{system} & = \sum_{c \in C, e \in E}P(c,e) \ CS_{transition}(e, c) \\ 
%             & = \sum_{c \in C, e \in E}P(c)P(e \mid c) \ CS_{transition}(e, c)
% \end{align*}

Our expectation is that, broadly, measures of causation should peak in their values when determinism is maximized and degeneracy is minimized. And indeed, that is what we find in the bipartite model across the measures of Section \ref{sec:causation_measures} and whether they are applied in a state-dependent / actual causation sense or in a global expectation sense (with the exception of the bit-flip measure, but this may be a function of our arbitrary state-labeling, since it is sensitive to that).

Furthermore, we consider different intervention distributions used to probe counterfactual space: the maximum entropy distribution where all states are equally and exhaustively probed; the stationary distribution where the states are weighted according to their frequency of occurrence in the long-term dynamic of the system; the local perturbation distribution, where a subset of the full state-space is probed by considering states that are close to the candidate cause according to some criteria of distance (e.g., Hamming distance).

The majority of the measures of causation increase with the determinism of the model and decrease as the model gets more degenerate (Figure \ref{fig:figure_3}). Moreover, the system level behavior of the causation measures, i.e. average across all state transitions, is dominated by that of the main transitions, which is consistent with the idea that these transitions concentrate the causal powers of the system. Note that these results are shown using local perturbations, but using the other intervention distributions led to qualitatively similar results the maximum-entropy distribution and the observational distribution (data shown for the causal primitives in Supplementary Section \ref{fig:figure_S2}). This indicates that local perturbations may indeed provide an efficient surrogate for computing causal powers without relying on the exhaustive exploration of counterfactual space or using an observational distribution that reflects the system's dynamics rather than its causal structure.

\section{\label{sec:macroscale_causation}Macroscale causation}

Causal emergence ($CE$) is computed as the difference between the macroscale's causal relationships and the microscale's causal relationships with respect to a given measure of causation.

$$ CE = CS_{macro} - CS_{micro} $$

If $CE$ is positive, there is causal emergence, i.e., the macroscale provides a better causal account of the system than the microscale. This can be interpreted as the macroscale doing more causal work, being more powerful, strong, or more informative, depending on how the chosen measure of causation is itself interpreted. A negative value indicates \textit{causal reduction}, which is when the microscale gives the superior causal account. Note that the theory is agnostic as to whether emergence or reduction occurs.

\subsection{\label{sec:macroscale_model}Modeling macroscales}

In order to calculate causal emergence, both a microscale and macroscale must be defined. It should be note that the theory is scale-relative, in that one starts with a microscale that is not necessarily some fundamental physical microscale. It is just some lower-bound scale. In neuroscience, for instance, this may be the scale of individual synapses. A macroscale is then some dimension reduction of the microscale, like a coarse-graining (an averaging) or black-boxing (a leaving of variables exogenous) or more generally just any summary statistic that recasts the system with less parameters \cite{klein_emergence_2020}. E.g., in the neurosciences a macroscale may be a local-field potential or neuronal population or even entire brain regions.

Previous research has laid out clear examples and definitions of macroscales in different system types \cite{hoel_quantifying_2013, hoel_when_2017, klein_emergence_2020}. One important note is that macroscales should be dynamically consistent with their underlying microscale. This means that the macroscale is not just derivable from the microscale (supervenience) but also that the macroscale behaves identically or similarly (in terms of its trajectory, dynamics, or state-transitions over time). Mathematical definitions of consistency between scales have been proposed \cite{klein_emergence_2020}; however, here we can eschew this issue as the macroscale for the bipartite model we use automatically ensures consistency by simply grouping each side of the bipartition. Specifically, we use a microscale with $N=16$ microstates $\Omega_{micro} = \Omega_A \bigcup \Omega_B = \{0000, 0001, 0010, 0011, 0100, 0101, 0110, 0111\} \bigcup \{1111, 1110, 1101, 1100, 1011, 1010, 1001, 1000\}$ and two macrostates $\Omega_{macro} = \{\textrm{ON}, \textrm{OFF}\}$ defined by the coarse-graining function $h : \Omega_{micro} \rightarrow \Omega_{macro}$, with $h(\Omega_A) = \textrm{ON}$ and $h(\Omega_B) = \textrm{OFF}$ (Figure \ref{fig:figure_2}B).

This coarse-grains the bipartite model into a simple two-state system at the macroscale, which trades off dynamically (a NOT gate with a self-loop). This macroscale is deterministic (each macrostate transitions solely to the other) and non-degenerate (each macrostate has only one possible cause). Note that, in our bipartite model, the macroscale is deterministic, non-degenerate, and dynamically consistent no matter the underlying noise in the microscale. This allows us to compare a consistent macroscale against parameterizations of noise, like increases in indeterminism and degeneracy. It's also worth noting that for the macroscale grouping of the bipartite model, the stationary intervention distribution, maximum entropy distribution, and local intervention distribution, are all identical at the macroscale, ensuring clear comparisons.

\subsection{\label{sec:causal_emergence}All measures of causation assessed show causal emergence}

\begin{figure}[h!]
\begin{center}
    \includegraphics[width=1\textwidth]{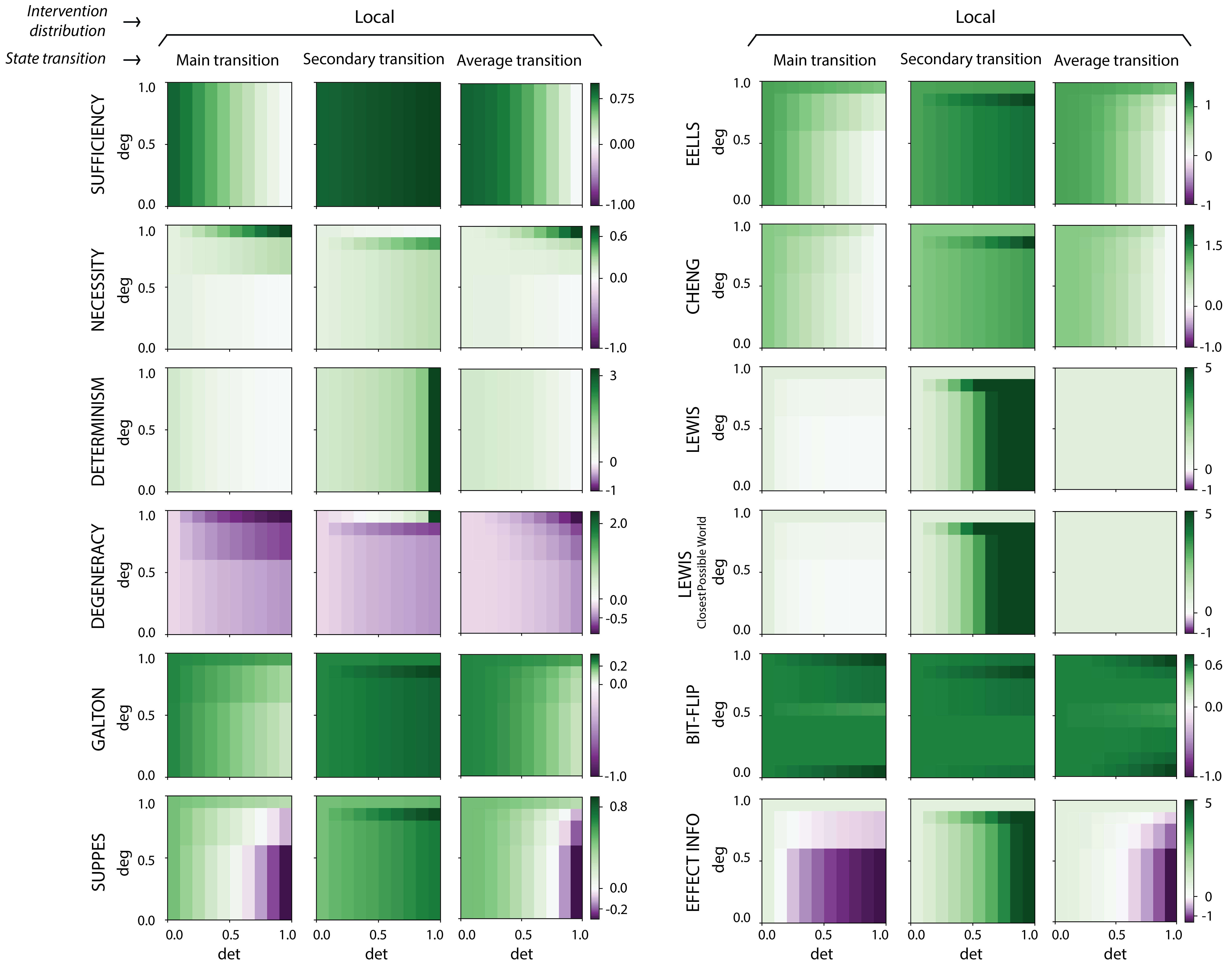}
  \caption{\textbf{Causal emergence is widespread across choice of measure of causation and intervention distribution.}
  Heatmaps of causal emergence (CE) and causal reduction (CR) is shown for all measures of causation and causal primitives computed in the bipartite Markov chain model. Causal emergence is calculated as the difference between the causation metric calculated in the macroscale and in the microscale, such that positive values (green) amount to CE and negative values (purple) to CR. CE/CR is assessed using a local intervention distribution, in which a subset of counterfactuals by perturbing the cause around "close" states. In each of the three columns, CE/CR is assessed over different state transitions of the system: a main transition with a strong causal strength ($0000 \rightarrow 1111$), a secondary transition with a weak causal strength ($0000 \rightarrow 1110$) and the expectation over all state transitions. The joint probability $P(c, e)$ used to compute the expectation is obtained using the transition probabilities $P(e \mid c)$ and the stationary intervention distribution $P_{obs}(c)$. For each measure (row), a common scale is used (shown in the colorbar). Causal emergence across the full combinations of intervention distributions and transitions can be found in Supplementary Figure \ref{fig:figure_S3}.}
  \label{fig:figure_4}

\end{center}
\end{figure}

\begin{figure}[ht]
\begin{center}
    \includegraphics[width=0.6\textwidth]{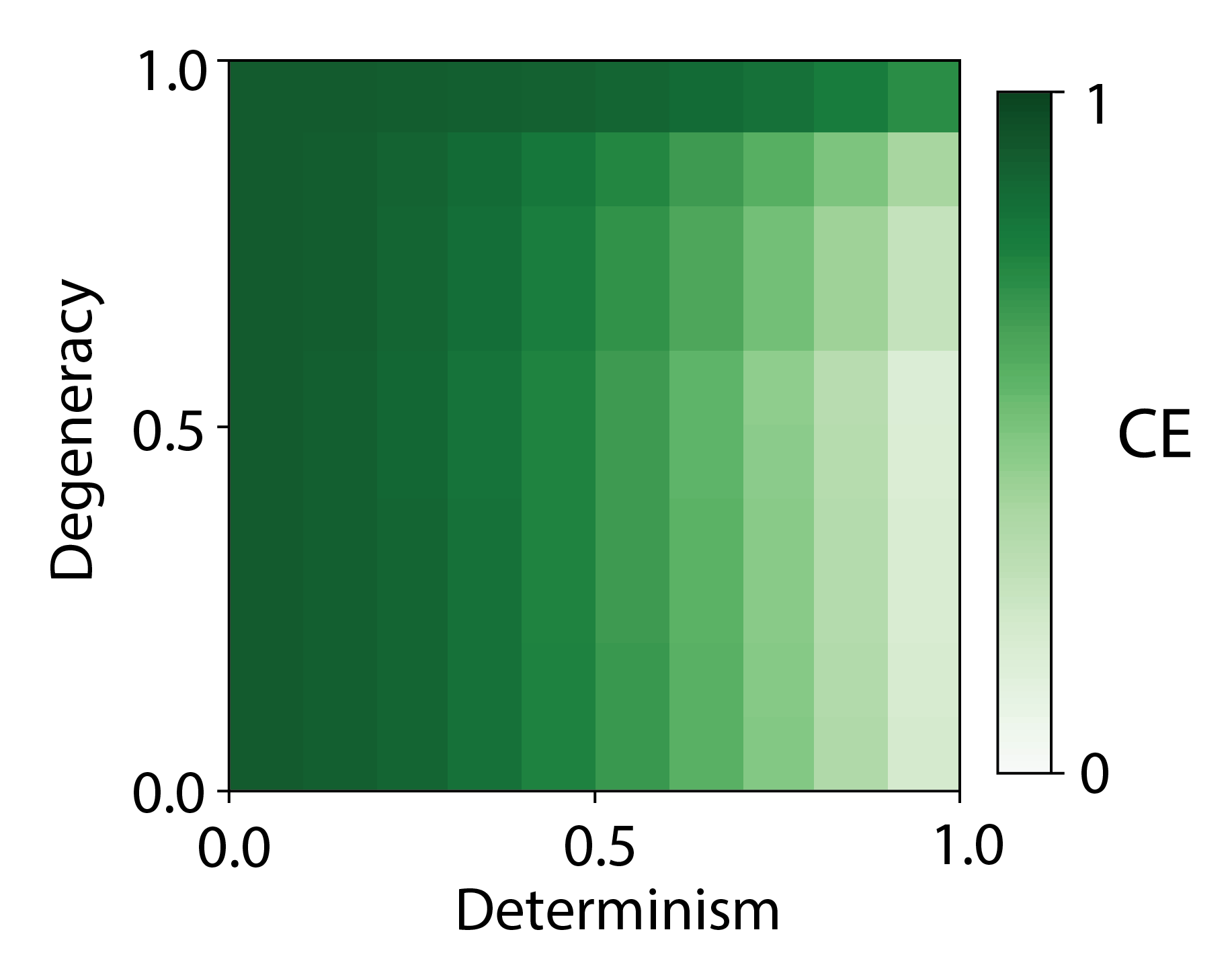}
  \caption{\textbf{Causal emergence occurs when the microscale is noisy.} Average behavior of causal emergence across the eight causation measures and four causal primitives for the bipartite Markov chain model. All twelve metrics were normalized to range from -1 to 1 by dividing each metric by its maximum absolute value of CE/CR and then combined through a simple average at each value of determinism and degeneracy. Values were all positive, ranging from low CE (light green) to high CE values (dark green) shown in the colorbar.}
  \label{fig:figure_5}
\end{center}
\end{figure}

Taking into consideration different transitions in the model and employing different intervention distributions, all measures of causation exhibited instances of causal emergence, as shown in Figure \ref{fig:figure_4}). This is likely because the causal primitives demonstrated causal emergence, and the measures are universally composed of or closely related to these primitives. Exactly as would be predicted by the idea that macroscales provide error-correction of noise in causal relationships, causal emergence is greater when determinism is low and degeneracy is high in the microscale across the set of measures (see Figure \ref{fig:figure_5}). Moreover, causal emergence occurred most prominently in secondary transitions, where causal strength in the microscale was shown to be generally lower due to noise, than in main transitions. There were even cases of "super causal emergence" wherein a microscale transition has a preventative role due to a negative value while the macroscale transition has a positive value, according to the same measure. Additionally, at the global system level, such as at the expectation of the measures of causation, there was also significant amount of causal emergence in certain system architecture domains (particularly those with more uncertainty).

Note that the ubiquity of causal emergence hinges on no particular way of performing the intervention distribution that all measures implicitly require be specified in their application. Causal emergence was present across measures of causation calculated using the maximum-entropy distribution, the observational intervention distribution, and the local intervention distribution (see Figure \ref{fig:figure_S3} in Supplementary Information), although distributed slightly differently depending on choice. Indeed, the only condition to not show causal emergence was the overall effective information when using the observational distribution. This was known \cite{hoel_when_2017} but was also pointed out by Scott Aaronson \cite{aaronson_higher-level_2017} as a possible criticism of the theory of causal emergence, since the mutual information (the effective information under the observational distribution) is not higher at a macroscale. First, as we show here, causal emergence still appears in the individual transition's effect information under the observational distribution, meaning that even under the observational distribution only in the average transition (not across all of them) was there no causal emergence in this condition (recall that the effective information is the average of the effect information). Additionally, the mutual information is not traditionally a causal measure \cite{cover_elements_2006}. Nor is it a monolithic quantity, but can be decomposed into synergistic, unique, and redundant information. Recent research has shown that the synergistic and unique mutual information can indeed increase at a macroscale, indicating that the non-redundant bits of the mutual information show causal emergence \cite{varley_emergence_2021}. Overall, in context of the results from other measures this indicates solely that effective information is a conservative measure of causal emergence, rather than a liberal one, compared to other measures of causation. 

Finally, in order to ensure that these results did not hinge on the symmetry of the bipartite model ($n_A=n_B$) we assessed causal emergence in an asymmetric bipartite models as well, which also showed causal emergence across measures (see Figure \ref{fig:figure_S3}).

\section{\label{sec:discussion}Discussion}

Causal emergence is when a measure of causation returns a higher value by having more causal strength, power, informativeness, predictiveness, or causal work (depending on the details of the measure of causation) at the macroscale vs. the microscale of a system. It is possible because macroscales can provide the advantage of noise minimization. That is, emergent scales are those that perform error-correction over their underlying microscale causal relationships. Indeed, we've shown that causal emergence is widespread across popular measures of causation with independent origins in diverse fields. This is because all of these measures are sensitive to noise in the form of indeterminism (uncertainty over the future) and degeneracy (uncertainty over the past). We refer to these terms, along with their simpler forms of sufficiency and necessity, as "causal primitives" since measures are either sensitive to them or even directly constructed from them. Notably, across the more than a dozen independent measures of causation we examined, all demonstrated causal emergence in a bipartite model system in conditions of high uncertainty over state transitions (low determinism, high degeneracy). This was true across a number of possible assumptions of how those measures are applied, showing the robustness of this theory of emergence.

The consilience of the measures examined here provide a bedrock for previous research which has already shown causal emergence using more complex information-theoretic measures of causation like effective information \cite{hoel_quantifying_2013}, the integrated information \cite{hoel_can_2016}, and also recently the synergistic information \cite{varley_emergence_2021}. Interestingly enough, we find that effective information, despite being the original measure proposed to capture causal emergence, is the most conservative in our sample.

One interesting discovery of this investigation is the similarities and agreements within measures of causation themselves. Broadly, we find that causation is not itself a primitive notion but can be decomposed along two dimensions (a finding in agreement with previous authors \cite{pearl_causality_2009, mackie_causes_1965}). These two dimensions are, in the philosophical literature, referred to as sufficiency and necessity; as we show, these are specific cases of determinism and degeneracy, respectively. Successful measures of causation are sensitive to both dimensions. Indeed, it is the sensitivity to these terms, and the uncertainty they capture, that guarantees the possibility of causal emergence of such measures.

It's worth noting that the measures of causation we examined need a space of possibilities, or counterfactuals, to be specified, in order to apply the measure. Here, we represent this choice mathematically using an intervention distribution. We find that causal emergence is relatively invariant across choice of intervention distribution, indicating that it is a robust phenomenon. While the choice of intervention distribution in the majority of measures doesn't affect the possibility of causal emergence, we advocate for our notion of "local interventions" as being a step forward for mathematical measures of causation, as it offers a compromise between a maximum-entropy approach (all possibilities considered) and a minimal-difference approach (only the closest possibility is considered).

Despite its ubiquity across measures and background conditions, the existence of emergence itself is not trivially guaranteed. Rather, it is a function of system architecture or dynamics. As we have shown, in dynamic domains of deterministic and time-reversible system mechanics, causal reduction dominates. However, in scientific models these conditions are quite rare, as science deals with mainly with open systems exposed to outside uncertainty or, alternatively, systems with inherent uncertainty. Even systems with irreducibly small amounts of noise can have that noise amplified into significant uncertainty after dynamical iteration \cite{strogatz_nonlinear_2014}. Therefore, we expect causal emergence to be common across the many scales and models of science.

The development of complex systems science was based on novel insights into how complexity can arise via iteration of simple rules \cite{wolfram_new_2002, newman_resource_2011, crutchfield_evolution_1995}; not only that, it was based around a family of measures of complexity \cite{gell-mann_what_1995}. The development of a science of emergence should be based on causal relationships (captured by the family of measures of causation) and the noise-minimizing properties of macroscales. Ultimately, this work provides a necessary toolkit for the scientific identification of emergent scales of function, along with optimal modeling choices, interventions, and explanations.

\bibliographystyle{unsrt}  
\bibliography{references}

\begin{thebibliography}{10}

\bibitem{pearl_causality_2009}
Judea Pearl.
\newblock {\em Causality}.
\newblock Cambridge University Press, Cambridge, 2 edition, 2009.

\bibitem{fitelson_probabilistic_2010}
Branden Fitelson and Christopher Hitchcock.
\newblock Probabilistic {Measures} of {Causal} {Strength}.
\newblock {\em Causality in the Sciences}, January 2010.

\bibitem{massimini_perturbational_2009}
Marcello Massimini, Melanie Boly, Adenauer Casali, Mario Rosanova, and Giulio
  Tononi.
\newblock A perturbational approach for evaluating the brain's capacity for
  consciousness.
\newblock In {\em Progress in {Brain} {Research}}, volume 177, pages 201--214.
  Elsevier, 2009.

\bibitem{chettih_single-neuron_2019}
Selmaan~N. Chettih and Christopher~D. Harvey.
\newblock Single-neuron perturbations reveal feature-specific competition in
  {V1}.
\newblock {\em Nature}, 567(7748):334--340, March 2019.
\newblock Bandiera\_abtest: a Cg\_type: Nature Research Journals Number: 7748
  Primary\_atype: Research Publisher: Nature Publishing Group Subject\_term:
  Computational neuroscience;Neural circuits;Sensory processing;Visual system
  Subject\_term\_id:
  computational-neuroscience;neural-circuit;sensory-processing;visual-system.

\bibitem{sporns_brain_2007}
Olaf Sporns.
\newblock Brain connectivity.
\newblock {\em Scholarpedia}, 2(10):4695, October 2007.

\bibitem{fingelkurts_functional_2005}
Andrew~A. Fingelkurts, Alexander~A. Fingelkurts, and Seppo Kähkönen.
\newblock Functional connectivity in the brain--{Is} it an elusive concept?
\newblock {\em Neuroscience and Biobehavioral Reviews}, 28(8):827--836, 2005.
\newblock Place: Netherlands Publisher: Elsevier Science.

\bibitem{gell-mann_what_1995}
Murray Gell-Mann.
\newblock What is complexity? {Remarks} on simplicity and complexity by the
  {Nobel} {Prize}-winning author of {The} {Quark} and the {Jaguar}.
\newblock {\em Complexity}, 1(1):16--19, 1995.
\newblock Publisher: John Wiley \& Sons, Ltd.

\bibitem{hoel_quantifying_2013}
Erik~P. Hoel, L.~Albantakis, and G.~Tononi.
\newblock Quantifying causal emergence shows that macro can beat micro.
\newblock {\em Proceedings of the National Academy of Sciences},
  110(49):19790--19795, December 2013.

\bibitem{chvykov_causal_2020}
Pavel Chvykov and Erik Hoel.
\newblock Causal {Geometry}.
\newblock {\em arXiv:2010.09390 [hep-th, physics:physics]}, October 2020.
\newblock arXiv: 2010.09390.

\bibitem{hoel_emergence_2020}
Erik Hoel and Michael Levin.
\newblock Emergence of informative higher scales in biological systems: a
  computational toolkit for optimal prediction and control.
\newblock {\em Communicative \& Integrative Biology}, 13(1):108--118, January
  2020.
\newblock Publisher: Taylor \& Francis \_eprint:
  https://doi.org/10.1080/19420889.2020.1802914.

\bibitem{klein_emergence_2020}
Brennan Klein and Erik Hoel.
\newblock The {Emergence} of {Informative} {Higher} {Scales} in {Complex}
  {Networks}.
\newblock {\em Complexity}, 2020:e8932526, April 2020.

\bibitem{varley_causal_2020}
Thomas~F. Varley.
\newblock Causal {Emergence} in {Discrete} and {Continuous} {Dynamical}
  {Systems}.
\newblock {\em arXiv:2003.13075 [nlin]}, March 2020.
\newblock arXiv: 2003.13075.

\bibitem{zhang_neural_2022}
Jiang Zhang.
\newblock Neural {Information} {Squeezer} for {Causal} {Emergence}.
\newblock {\em arXiv:2201.10154 [physics]}, January 2022.
\newblock arXiv: 2201.10154.

\bibitem{klein_evolution_2021}
Brennan Klein, Erik Hoel, Anshuman Swain, Ross Griebenow, and Michael Levin.
\newblock Evolution and emergence: higher order information structure in
  protein interactomes across the tree of life.
\newblock {\em Integrative Biology}, 13(12):283--294, 2021.

\bibitem{yuste_neuron_2015}
Rafael Yuste.
\newblock From the neuron doctrine to neural networks.
\newblock {\em Nature Reviews Neuroscience}, 16(8):487--497, August 2015.
\newblock Bandiera\_abtest: a Cg\_type: Nature Research Journals Number: 8
  Primary\_atype: Reviews Publisher: Nature Publishing Group Subject\_term:
  Electrophysiology;Network models;Neural circuits Subject\_term\_id:
  electrophysiology;network-models;neural-circuit.

\bibitem{buxhoeveden_minicolumn_2002}
Daniel~P. Buxhoeveden and Manuel~F. Casanova.
\newblock The minicolumn hypothesis in neuroscience.
\newblock {\em Brain}, 125(5):935--951, 2002.

\bibitem{yeo_organization_2011}
B.~T.~Thomas Yeo, Fenna~M. Krienen, Jorge Sepulcre, Mert~R. Sabuncu, Danial
  Lashkari, Marisa Hollinshead, Joshua~L. Roffman, Jordan~W. Smoller, Lilla
  Zöllei, Jonathan~R. Polimeni, Bruce Fischl, Hesheng Liu, and Randy~L.
  Buckner.
\newblock The organization of the human cerebral cortex estimated by intrinsic
  functional connectivity.
\newblock {\em Journal of Neurophysiology}, 106(3):1125--1165, September 2011.

\bibitem{hoel_can_2016}
Erik~P. Hoel, Larissa Albantakis, William Marshall, and Giulio Tononi.
\newblock Can the macro beat the micro? {Integrated} information across
  spatiotemporal scales.
\newblock {\em Neuroscience of Consciousness}, 2016(1):niw012, 2016.

\bibitem{chang_information_2020}
Acer Y.~C. Chang, Martin Biehl, Yen Yu, and Ryota Kanai.
\newblock Information {Closure} {Theory} of {Consciousness}.
\newblock {\em Frontiers in Psychology}, 11:1504, July 2020.
\newblock arXiv: 1909.13045.

\bibitem{hoel_when_2017}
Erik Hoel.
\newblock When the {Map} {Is} {Better} {Than} the {Territory}.
\newblock {\em Entropy}, 19(5):188, April 2017.

\bibitem{marshall_black-boxing_2018}
William Marshall, Larissa Albantakis, and Giulio Tononi.
\newblock Black-boxing and cause-effect power.
\newblock {\em PLOS Computational Biology}, 14(4):e1006114, April 2018.

\bibitem{aaronson_higher-level_2017}
Scott Aaronson.
\newblock Higher-level causation exists (but {I} wish it didn’t), June 2017.

\bibitem{dewhurst_causal_2021}
Joe Dewhurst.
\newblock Causal emergence from effective information: {Neither} causal nor
  emergent?
\newblock {\em Thought: A Journal of Philosophy}, 10(3):158--168, 2021.
\newblock Publisher: John Wiley \& Sons, Ltd.

\bibitem{tegmark_improved_2016}
Max Tegmark.
\newblock Improved {Measures} of {Integrated} {Information}.
\newblock {\em PLOS Computational Biology}, 12(11):e1005123, November 2016.
\newblock Publisher: Public Library of Science.

\bibitem{mediano_beyond_2019}
Pedro A.~M. Mediano, Fernando Rosas, Robin~L. Carhart-Harris, Anil~K. Seth, and
  Adam~B. Barrett.
\newblock Beyond integrated information: {A} taxonomy of information dynamics
  phenomena.
\newblock {\em arXiv:1909.02297 [physics, q-bio]}, September 2019.
\newblock arXiv: 1909.02297.

\bibitem{bayne_axiomatic_2018}
Tim Bayne.
\newblock On the axiomatic foundations of the integrated information theory of
  consciousness.
\newblock {\em Neuroscience of Consciousness}, 2018(1), January 2018.

\bibitem{varley_emergence_2021}
Thomas Varley and Erik Hoel.
\newblock Emergence as the conversion of information: {A} unifying theory.
\newblock {\em arXiv:2104.13368}, April 2021.
\newblock arXiv: 2104.13368.

\bibitem{mediano_greater_2021}
Pedro A.~M. Mediano, Fernando~E. Rosas, Andrea~I. Luppi, Henrik~J. Jensen,
  Anil~K. Seth, Adam~B. Barrett, Robin~L. Carhart-Harris, and Daniel Bor.
\newblock Greater than the parts: {A} review of the information decomposition
  approach to causal emergence.
\newblock {\em arXiv:2111.06518 [nlin, q-bio]}, November 2021.
\newblock arXiv: 2111.06518.

\bibitem{rosas_reconciling_2020}
Fernando~E. Rosas, Pedro A.~M. Mediano, Henrik~J. Jensen, Anil~K. Seth, Adam~B.
  Barrett, Robin~L. Carhart-Harris, and Daniel Bor.
\newblock Reconciling emergences: {An} information-theoretic approach to
  identify causal emergence in multivariate data.
\newblock {\em PLOS Computational Biology}, 16(12):e1008289, December 2020.
\newblock Publisher: Public Library of Science.

\bibitem{nadathur_causal_2020}
Prerna Nadathur and Sven Lauer.
\newblock Causal necessity, causal sufficiency, and the implications of
  causative verbs.
\newblock {\em Glossa: a journal of general linguistics}, 5(1):49, June 2020.
\newblock Number: 1 Publisher: Ubiquity Press.

\bibitem{hume_enquiry_1748}
David Hume.
\newblock {\em An {Enquiry} concerning {Human} {Understanding}}.
\newblock 1748.

\bibitem{illari_causality_2014}
Phyllis Illari and Federica Russo.
\newblock {\em Causality: {Philosophical} {Theory} meets {Scientific}
  {Practice}}.
\newblock Oxford University Press, Oxford, New York, December 2014.

\bibitem{eells_probabilistic_1991}
Ellery Eells.
\newblock {\em Probabilistic {Causality}}.
\newblock Cambridge University Press, 1991.

\bibitem{albantakis_what_2019}
Larissa Albantakis, William Marshall, Erik Hoel, and Giulio Tononi.
\newblock What {Caused} {What}? {A} {Quantitative} {Account} of {Actual}
  {Causation} {Using} {Dynamical} {Causal} {Networks}.
\newblock {\em Entropy}, 21(5):459, May 2019.

\bibitem{suppes_probabilistic_1968}
Patrick Suppes.
\newblock {\em A {Probabilistic} {Theory} of {Causality}}.
\newblock Amsterdam: North-Holland Pub. Co., 1968.

\bibitem{hitchcock_probabilistic_2018}
Christopher Hitchcock.
\newblock Probabilistic {Causation}.
\newblock In Edward~N. Zalta, editor, {\em The {Stanford} {Encyclopedia} of
  {Philosophy}}. Metaphysics Research Lab, Stanford University, spring 2021
  edition, 2018.

\bibitem{cheng_causes_1991}
Patricia~W. Cheng and Laura~R. Novick.
\newblock Causes versus enabling conditions.
\newblock {\em Cognition}, 40(1):83--120, August 1991.

\bibitem{lewis_causation_1973}
David Lewis.
\newblock Causation.
\newblock {\em Journal of Philosophy}, 70(17):556--567, 1973.

\bibitem{lewis_postscripts_1986}
David Lewis.
\newblock Postscripts to '{Causation}'.
\newblock {\em Philosophical Papers Vol. Ii}, 1986.

\bibitem{floridi_information_2010}
Luciano Floridi.
\newblock Information, possible worlds and the cooptation of scepticism.
\newblock {\em Synthese}, 175:63--88, 2010.
\newblock Publisher: Springer.

\bibitem{daniels_criticality_2018}
Bryan~C. Daniels, Hyunju Kim, Douglas Moore, Siyu Zhou, Harrison~B. Smith,
  Bradley Karas, Stuart~A. Kauffman, and Sara~I. Walker.
\newblock Criticality {Distinguishes} the {Ensemble} of {Biological}
  {Regulatory} {Networks}.
\newblock {\em Physical Review Letters}, 121(13):138102, September 2018.
\newblock Publisher: American Physical Society.

\bibitem{tononi_measuring_2003}
Giulio Tononi and Olaf Sporns.
\newblock Measuring information integration.
\newblock {\em BMC Neuroscience}, page~20, 2003.

\bibitem{griffiths_measuring_2015}
Paul~E. Griffiths, Arnaud Pocheville, Brett Calcott, Karola Stotz, Hyunju Kim,
  and Rob Knight.
\newblock Measuring {Causal} {Specificity}.
\newblock {\em Philosophy of Science}, 82(4):529--555, 2015.
\newblock Publisher: The University of Chicago Press.

\bibitem{mackie_causes_1965}
J.~L. Mackie.
\newblock Causes and {Conditions}.
\newblock {\em American Philosophical Quarterly}, 2(4):245--264, 1965.
\newblock Publisher: University of Illinois Press.

\bibitem{pearl_book_2017}
Judea Pearl and Dana Mackenzie.
\newblock {\em The {Book} of {Why}: {The} {New} {Science} of {Cause} and
  {Effect}}.
\newblock Basic Books, New York, 1° edizione edition, 2017.

\bibitem{oizumi_phenomenology_2014}
Masafumi Oizumi, Larissa Albantakis, and Giulio Tononi.
\newblock From the {Phenomenology} to the {Mechanisms} of {Consciousness}:
  {Integrated} {Information} {Theory} 3.0.
\newblock {\em PLoS Computational Biology}, 10(5):e1003588, May 2014.

\bibitem{balduzzi_integrated_2008}
David Balduzzi and Giulio Tononi.
\newblock Integrated {Information} in {Discrete} {Dynamical} {Systems}:
  {Motivation} and {Theoretical} {Framework}.
\newblock {\em PLoS Computational Biology}, 4(6):e1000091, June 2008.

\bibitem{halpern_actual_2016}
Joseph~Y. Halpern.
\newblock {\em Actual {Causality}}.
\newblock MIT Press, Cambridge, MA, USA, August 2016.

\bibitem{juel_when_2019}
Bjørn~Erik Juel, Renzo Comolatti, Giulio Tononi, and Larissa Albantakis.
\newblock When is an action caused from within? {Quantifying} the causal chain
  leading to actions in simulated agents.
\newblock pages 477--484. MIT Press, July 2019.

\bibitem{adami_use_2012}
Christoph Adami.
\newblock The use of information theory in evolutionary biology: {Information}
  theory in evolutionary biology.
\newblock {\em Annals of the New York Academy of Sciences}, 1256(1):49--65, May
  2012.

\bibitem{timme_tutorial_2018}
Nicholas~M. Timme and Christopher Lapish.
\newblock A {Tutorial} for {Information} {Theory} in {Neuroscience}.
\newblock {\em eNeuro}, 5(3):ENEURO.0052--18.2018, September 2018.

\bibitem{cover_elements_2006}
Thomas~M. Cover and Joy~A. Thomas.
\newblock {\em Elements of {Information} {Theory} 2nd {Edition}}.
\newblock Wiley-Interscience, Hoboken, N.J, 2nd edition edition, July 2006.

\bibitem{strogatz_nonlinear_2014}
Steven Strogatz.
\newblock {\em Nonlinear {Dynamics} and {Chaos}, 2nd {Edition}: {With}
  {Applications} to {Physics}, {Biology}, {Chemistry}, and {Engineering}:
  {With} {Applications} to {Physics}, {Biology}, {Chemistry}, and
  {Engineering}, {Second} {Edition}}.
\newblock Westview Press, Boulder, CO, 2° edizione edition, 2014.

\bibitem{wolfram_new_2002}
Stephen Wolfram.
\newblock {\em A {New} {Kind} of {Science}}.
\newblock Wolfram Media Inc, Champaign, IL, first edition edition, 2002.

\bibitem{newman_resource_2011}
M.~E.~J. Newman.
\newblock Resource {Letter} {CS}–1: {Complex} {Systems}.
\newblock {\em American Journal of Physics}, 79(8):800--810, 2011.
\newblock Publisher: American Association of Physics Teachers.

\bibitem{crutchfield_evolution_1995}
J.~P. Crutchfield and M.~Mitchell.
\newblock The evolution of emergent computation.
\newblock {\em Proceedings of the National Academy of Sciences},
  92(23):10742--10746, November 1995.
\newblock Publisher: National Academy of Sciences Section: Research Article.

\end{thebibliography}

\section{\label{sec:SI}Supplementary Information}

\renewcommand{\thefigure}{S\arabic{figure}}
\setcounter{figure}{0}

\subsection{\label{sec:SI_model}Parameterizing determinism and degeneracy in the bipartite model}

What follows is the detailed description of how we algorithmically vary $det$ and $deg$ in the bipartite Markov chain model. First we will label the $2^N=n$ states with binary strings and divide them into two groups $A$ and $B$ of size $n_A$ and $n_B$, respectively. For now, let us consider the symmetric case where $n_A=n_B$. For example, with $N=3$ we have states $\Omega = \Omega_A \cup \Omega_B = \{000, 001, 010, 011\} \cup \{111, 110, 101, 100\}$) (Figure \ref{fig:figure_2}A, top). The model's dynamics is governed by a transition probability matrix, where for a given state $c\in \Omega$ the system is in, it can transition to a state $e\in \Omega$ with probability given by $P(e \mid  c)$, such that any state transition defines a cause and effect pair (Figure \ref{fig:figure_2}A, bottom). Each cause state is paired to a main effect state in the opposite grouping through a mapping $f_A: \Omega_A \rightarrow \Omega_B$. A complementary mapping is given $f_B(s) = f_A(s\uparrow)\uparrow$, where $\uparrow$ is the state obtained by inverting all bits (e.g. $\uparrow 100 = 011$) (this is equivalent to reflecting the arrows in Figure \ref{fig:figure_2}A along the vertical axis). For a given state $c\in\Omega_A$ we have:

\begin{equation}
    P(e \mid  c) = \begin{cases}
    p, &\text{if $e = f_A(c)$ and $e \in \Omega_B$}\\
    \frac{(1 - p)}{N_B} &\text{if $e \neq f_A(c)$ and $e \in \Omega_B$}\\
    0 &\text{otherwise}
    \end{cases}
    \end{equation}

If $c\in\Omega_B$, we simply interchange $B$ for $A$ in the definition above. $0 \leq p \leq 1$ is the parameter which controls the determinism of the system, by concentrating or diluting the probability over the main effect (vs. the secondary effects) of a given cause. Essentially, we are simply narrowing or widening the "scope" of possible effects from a given state in order to increase or decrease the determinism, respectively.

\begin{figure}[ht]
\begin{center}
    \includegraphics[width=0.9\textwidth]{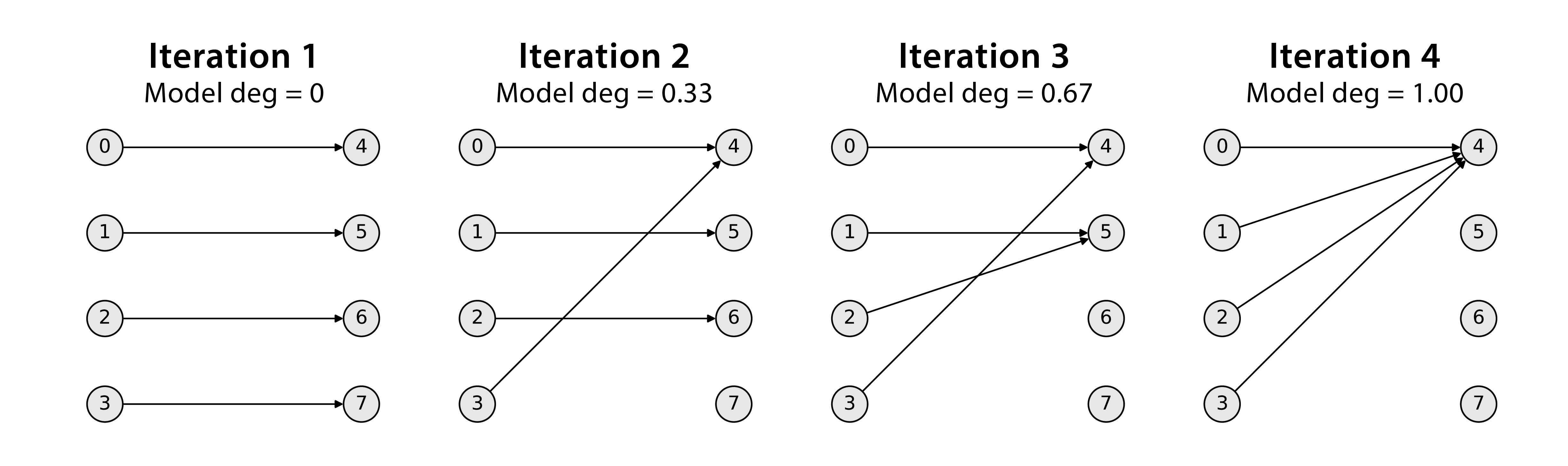}
    \caption{\textbf{Visualization of steps in the algorithm to increase degeneracy in the bipartite model.} Iterations of the algorithm for the bipartite Markov chain state-space with $n=8$ states. States are labeled from 0 to $n-1$, with states in the left column belonging to group $A$ and on the right to group $B$. The function $f_A: \Omega_A \rightarrow \Omega_B$ maps states in group $A$ to states in group $B$, associating every candidate cause $c\in \Omega_A$ to a main effect $f_A(c)=e^{*}$. At each iteration, a main effect is lost (i.e. the image $f(A)\in\Omega_B$ loses an element) as an arrow is moved an effect with the least number of arrows (with at least one arrow). The main effects of each cause progressively overlap until they are over a single state in group $B$. For $f_B$ the same algorithm is applied, but with the states $A$ and $B$ reversed.}
    \label{fig:figure_S1}
\end{center}
\end{figure}

In order to parametrically vary the degeneracy of the model we change the mapping $f_A$ (and its complement $f_B$), going from zero degeneracy where $f_A$ is injective, (maps different cause to different main effects according to $f_A(s) = \uparrow s$ to max degeneracy), where all causes in one group map to a single effect in the other group (Figure \ref{fig:figure_2}B). To increase the degeneracy in a step wise manner we use the following algorithm: we chose the "poorest" effect, i.e. the one with the least number of main causes ($argmin \mid f^{-1}(e)\mid$) but with at least one main cause, and move all its main causes to the next poorest effect. In this way, we progressively re-wire the system until all causes map to only one effect and maximal degeneracy is achieved. Essentially, we are simply moving main effects on top of one another sequentially---this increases the degeneracy (and decreases the necessity). A visual example of this can be see in Figure \ref{fig:figure_S1}.

However, it should be noted that our algorithmic methods for varying the determinism and degeneracy do not automatically ensure that it is changing the causal primitives as expected. This is because the algorithmic way of varying determinism and degeneracy (by varying the probabilities between main effects and second effects, and stacking main effects on top of targets, respectively) does not match one-to-one with the underlying mathematical properties of determinism and degeneracy. This is because there is no way to smoothly vary the actual mathematical properties in a simple algorithmic manner. 

\begin{figure}[ht]
\begin{center}
    \includegraphics[width=0.9\textwidth]{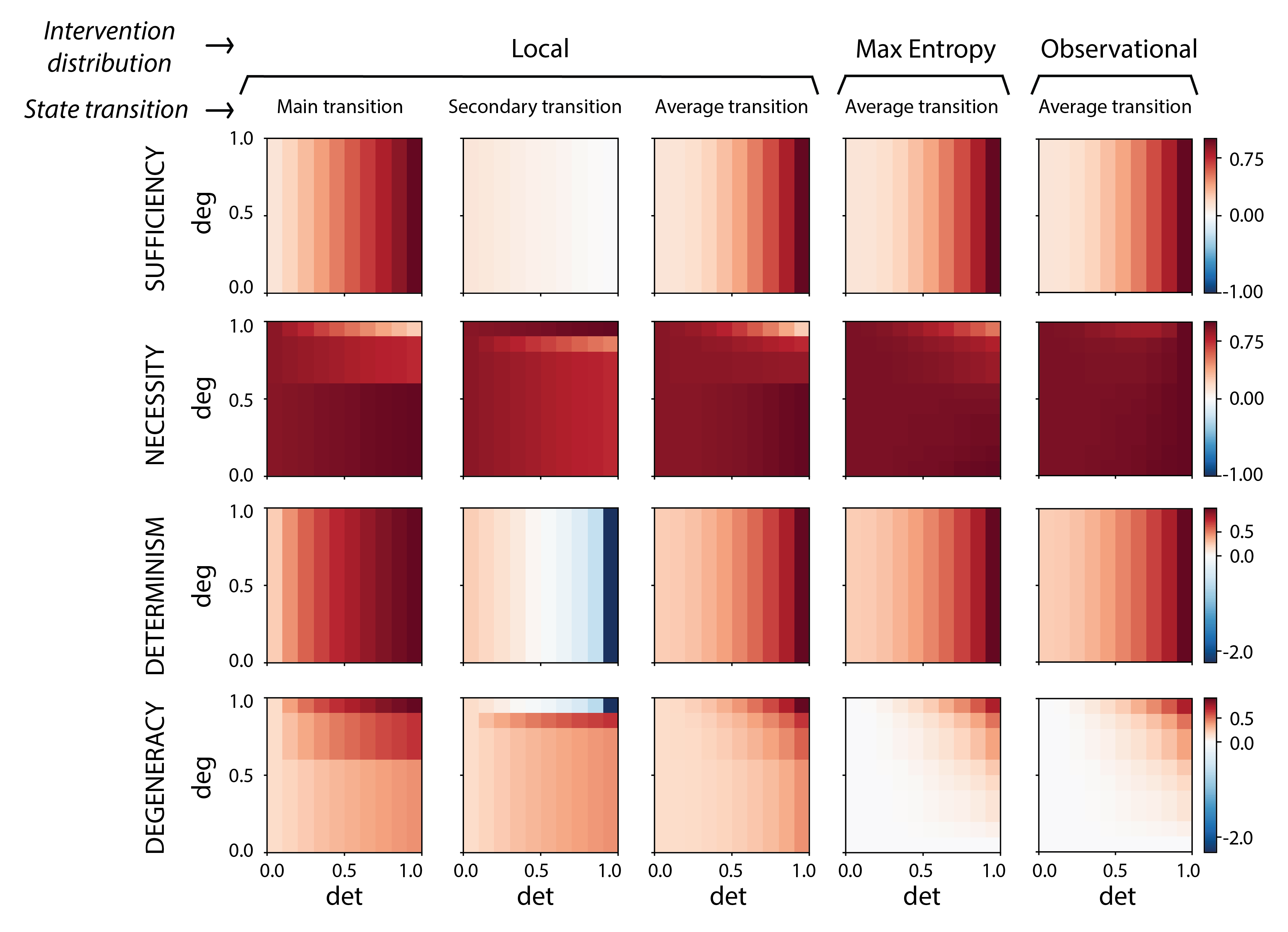}
  \caption{\textbf{Behavior of the causal primitives in the model system.}  Shown along the rows are the heatmaps of the causal primitives, i.e. suffiency, necessity, determinism and degeneracy, for different values of the model's degeneracy and determinism parameters. In the first three columns, the max entropy distribution is used to calculate the causal primitives and average the across transitions. In the first two columns, the causal primitives assessed for single state transition between a cause and an effect: first, between a cause and its main effect ($0000 \rightarrow 1111$), generally a strong causal link, and second, between a cause and a non principal effect, which generally we would expect to have a weaker causal strength ($0001 \rightarrow 1111$). In the third column, the simple average of the causal primitives across all the state transitions. In the fourth column the average of the causal primitives is shown, but using the stationary distribution to estimate the primitives and also compute the average. In the last column, the causal primitives average across all transitions computed using local perturbations.}
  \label{fig:figure_S2}
\end{center}
\end{figure}

However, when the causal primitives are computed over different parameters of the model and we considered their global behavior averaged across all transitions in the state-space, the algorithmic determinism and sufficiency indeed scale with their mathematical counterparts. Similarly, the degeneracy primitive scales proportionally to the model's degeneracy parameter, while necessity does so inversely (Figure \ref{fig:figure_S2}). Note that while determinism and sufficiency are independent of the model's degeneracy parameter, degeneracy and necessity are sensitive to the model's determinism parameter, simply due to its algorithmic construction. These results validate the bipartite model capacity to explore the behavior of the causal measures for different combinations of causal primitives, as modulated by the model's degeneracy and determinism parameter, although, due to the inability to vary degeneracy without varying the determinism, it does not do so over a perfectly symmetric manifold.

It's also interesting to note how the causal primitives behave differently for specific transitions with strong and weak causal link. The main transitions exhibit high sufficiency, determinism and necessity and low degeneracy (Figure \ref{fig:figure_S2}, first column), in particular, at regions of high determinism and low degeneracy of the parameter space of the model. This behavior dominates and appears at the level of the average quantities across all transitions (Figure \ref{fig:figure_S2}, last three columns). The secondary transitions show lower values in general for all causal primitives, coherent with the notion that they are endowed with weaker causal powers. In line with this, the determinism and degeneracy of the secondary transitions vary in the opposite manner of a main transition, peaking when the determinism of the model is low, and the degeneracy is high (Figure \ref{fig:figure_S2}, second column).

\begin{figure}[ht]
\begin{center}
    \includegraphics[width=1\textwidth]{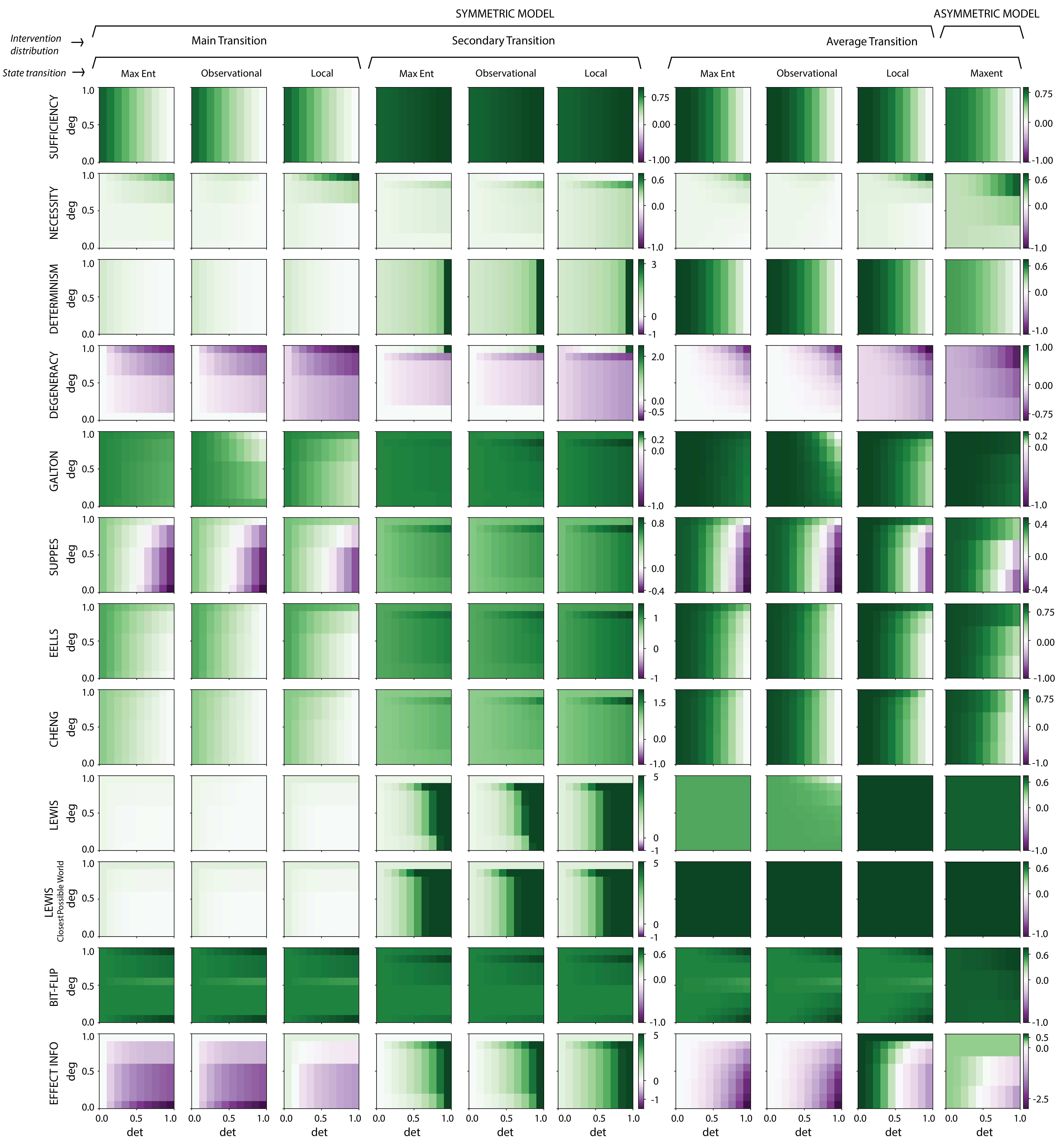}
    \caption{\textbf{Causal emergence is generally invariant to intervention distribution choice, as well as symmetry breaking.} An expanded version of Figure \ref{fig:figure_4} to include three different choices of intervention distributions, as well as what happens when the bipartite model is not perfectly symmetric. Heatmaps of causal emergence (CE) and causal reduction (CR) are shown for all measures of causation and causal primitives computed in the bipartite Markov chain model. Causal emergence is calculated as the difference between the causation metric calculated in the macroscale and in the microscale, such that positive values (green) amount to CE and negative values (purple) to CR. CE/CR is assessed using a maximum entropy distribution, a local intervention distribution, and the observational distribution. CE/CR is also assessed over different state transitions of the system: a main transition with a strong causal strength ($0000 \rightarrow 1111$), a secondary transition with a weak causal strength ($0000 \rightarrow 1110$) and the expectation over all state transitions. The joint probability $P(c, e)$ used to compute the expectation using the observational intervention distribution $P_{obs}(C)$. However, the expectation of causal emergence is relatively invariant across even this choice as well (data not shown). For each measure (row), a common scale is used (shown in the colorbar). In the last column, the model was calculated using an asymmetric version of the bipartite model with $n_A=13$ and $n_B=3$ states on each macro group, instead of $n_A=n_B=8$ used in the rest of the paper.}
    \label{fig:figure_S3}
\end{center}
\end{figure}

\end{document}